\documentclass{article}

\usepackage{arxiv}

\usepackage[utf8]{inputenc} 
\usepackage[T1]{fontenc}    
\usepackage{hyperref}       
\usepackage{url}            
\usepackage{booktabs}       
\usepackage{amsfonts}       
\usepackage{nicefrac}       
\usepackage{microtype}      
\usepackage{lipsum}
\usepackage{fancyhdr}       
\usepackage{graphicx}       
\graphicspath{{figure/}}     
\usepackage{amsmath,amssymb,amsfonts}
\usepackage{multirow}
\usepackage{xcolor}

\newcommand{\quotes}[1]{``#1''}

\usepackage{caption}
\usepackage{subcaption}
\usepackage{tabularray}

\title{WISDOM: An AI-powered framework for emerging research detection using \underline{w}eak s\underline{i}gnal analysi\underline{s} and advance\underline{d} t\underline{o}pic \underline{m}odeling}

\author{
  Ashkan Ebadi \\
  Digital Technologies Research Centre \\
  National Research Council Canada \\
  Toronto, ON M5T 3J1, Canada\\
  \texttt{ashkan.ebadi@nrc-cnrc.gc.ca} \\
  \And
  Alain Auger \\
  Science and Technology Foresight and Risk Assessment Unit \\
  Defence Research and Development Canada \\
  Quebec City, QC G3J 1X5, Canada\\
  \And
  Yvan Gauthier \\
  Digital Technologies Research Centre \\
  National Research Council Canada \\
  Ottawa, ON K1A 0R6, Canada \\
}

\begin{document}
\maketitle

\begin{abstract}
The landscape of science and technology is characterized by its dynamic and evolving nature, constantly reshaped by new discoveries, innovations, and paradigm shifts. Moreover, science is undergoing a remarkable shift towards increasing interdisciplinary collaboration, where the convergence of diverse fields fosters innovative solutions to complex problems. Detecting emerging scientific topics is paramount as it enables industries, policymakers, and innovators to adapt their strategies, investments, and regulations proactively. As the common approach for detecting emerging technologies, despite being useful, bibliometric analyses may suffer from oversimplification and/or misinterpretation of complex interdisciplinary trends. In addition, relying solely on domain experts to pinpoint emerging technologies from science and technology trends might restrict the ability to systematically analyze extensive information and introduce subjective judgments into the interpretations. To overcome these drawbacks, in this work, we present an automated artificial intelligence-enabled framework, called WISDOM, for detecting emerging research themes using advanced topic modeling and weak signal analysis. The proposed approach can assist strategic planners and domain experts in more effectively recognizing and tracking trends related to emerging topics by swiftly processing and analyzing vast volumes of data, uncovering hidden cross-disciplinary patterns, and offering unbiased insights, thereby enhancing the efficiency and objectivity of the detection process. As the case technology, we assess WISDOM's performance in identifying emerging research as well as their trends, in the field of underwater sensing technologies using scientific papers published between 2004 and 2021.  
\end{abstract}


\section{Introduction}
\label{sec:intro}
Technologies have a substantial influence on individuals and society, reshaping our lives, ways of thinking, and behaviors, while also revolutionizing the modes of communication and knowledge acquisition \cite{kenski_tecnologias_2013}. For many years, researchers have been concerned with identifying new scientific/innovative areas and the emergence of novel technologies \cite{viet_analyzing_2021}. Moreover, early prediction of upcoming changes in the technological landscape is crucial for governments and businesses to strategically plan their research and development (R\&D) efforts to maintain innovation leadership and competitive advantage \cite{viet_analyzing_2021}. 

In recent decades, technology has been advancing very quickly. The high complexity of modern science along with its increasing interdisciplinarity \cite{viet_analyzing_2021}, have made technology forecast and strategic planning more challenging \cite{muhlroth_systematic_2018}. Scientometrics/bibliometrics, performed and/or interpreted by domain experts, is the common approach for detecting emerging technologies. Several scientometrics studies have explored diverse data sources, including scholarly articles and patents, utilizing varied methodologies to detect the rise and development of emerging technologies \cite{noh_identifying_2016,dotsika_identifying_2017}. Despite being useful, especially for exploratory analyses, these conventional methods come with certain drawbacks, e.g., relying on single-aspect indicators, requiring extensive manual involvement/intervention, and facing issues of subjectivity due to heavy dependence on domain expert input \cite{ebadi_detecting_2022}. With the recent advances in big digital data and advanced data analysis techniques, new opportunities arise for predicting technological trends and identifying emerging technologies \cite{ebadi_detecting_2022}. These approaches leverage substantial digital data volumes and the latest advancements in computer science.

Research focusing on emerging technologies has a considerable impact on competitiveness which in the realm of emerging technologies, can lead to a dominant position in the market. Owing to the rapid advancement of technologies, establishing an effective framework is imperative for streamlining choices concerning future scientific research and development endeavors. It is of paramount importance to offer evidence-based insights into forthcoming research trends and the emergence of novel technologies \cite{zamani_developing_2022}. This study endeavors to overcome these constraints and harness the capabilities of advanced analytical methods. We achieve this by introducing a multi-layer framework, called WISDOM, driven by artificial intelligence (AI), which combines diverse approaches such as natural language processing (NLP), topic modeling, and weak signal analysis to detect emerging research. To assess the WISDOM's efficacy in recognizing scientific trends and capturing initial cues, we concentrate on publications within the realm of underwater sensing spanning from 2004 to 2021, serving as a test case. 

We make at least six key contributions: 1) we use an advanced topic modeling technique, i.e., BERTopic \cite{grootendorst_bertopic_2022}, to identify the main research topics and trace their evolution. BERTopic utilizes transformers and a class-based variation of TF-IDF to establish dense clusters, facilitating the creation of high-quality topics that are easily interpretable while retaining crucial terms within the topic descriptions. 2) We incorporate contemporary principles of weak signal analysis into the process of identifying the emergence of new research topics. The benefit of combining weak signal analysis with BERTopic is that it enables the detection of subtle and emerging research trends or technologies that might not be immediately apparent through conventional methods, with no/limited manual intervention required. Weak signal analysis helps uncover nascent research developments that have the potential to significantly impact a particular scientific domain, allowing stakeholders to proactively adapt, invest, or strategize based on these insights. 3) We provide a versatile, modular, and multi-layered approach that can be adapted for use in various technology domains. The approach not only detects signals of emerging technologies it can also trace their evolution as well as patterns of interest over time. WISDOM facilitates quicker, more extensive, and quantitative analyses on a large scale, which can serve as a valuable decision-support tool for experts. 4) This work covers a span of nearly two decades (2004-2021), allowing for temporal analysis of emerging research trends. This temporal dimension adds depth to the insights, helping stakeholders understand how research themes have evolved over time. 5) The AI-enabled approach ensures that the detection process remains objective and unbiased. It eliminates human biases that may influence trend recognition, enhancing the reliability of the insights provided. And, 6) We demonstrate the practicality and effectiveness of our proposed framework by applying it to the specific domain of underwater sensing technologies. This case study serves as a real-world example of the framework's utility and performance. However, our framework's applicability extends beyond the case study presented here.

The structure of the paper unfolds as follows: Section~\ref{sec:rel_work} provides a review of prior studies. The data and methodology are detailed in Section~\ref{sec:data_methods}, followed by the presentation of our findings in Section~\ref{sec:results} and their subsequent discussion in Section~\ref{sec:discussion}. In Section~\ref{sec:lim}, we address the limitations of this research and outline potential avenues for future investigation.

\section{Related Work}
\label{sec:rel_work}
The rise of globalization and increasing competition has sparked a growing research interest in emerging technologies. In recent years, the number of publications centred around emerging technologies has been increasing \cite{zamani_developing_2022}. Emerging technologies have not only been debated in academic research, but they have also been a central topic in policy discussions and initiatives \cite{rotolo_what_2015}. However, diverse interpretations of what constitutes \quotes{\textit{emerging technologies}} exist, with certain definitions converging on common aspects while also highlighting varying dimensions and characteristics of this phenomenon. For instance, some definitions underscore the potential socio-economic impact that emerging technologies can wield (e.g., \cite{porter_measuring_2002}). Conversely, other definitions place significant emphasis on the uncertainty inherent in the emergence process (e.g., \cite{boon_exploring_2008}) or on the distinctive attributes of novelty and growth (e.g., \cite{small_identifying_2014}). 

In an important and comprehensive study, Rotolo et al. \cite{rotolo_what_2015} defined emerging technologies by five key attributes: radical novelty, fast growth, coherence, significant impact, and a degree of uncertainty and ambiguity. In particular, they identify an emerging technology as a technology that is profoundly innovative and exhibits relatively rapid growth. It is also marked by a certain level of consistency over time and possesses the potential to significantly influence socio-economic domains. However, its most notable impact is anticipated to occur in the future, and during the emergence phase, where there remains an element of uncertainty and ambiguity \cite{rotolo_what_2015}. In a broader definition, \cite{veletsianos_emergence_2016} tagged emerging technologies as those that have not gained widespread acceptance but are anticipated to have a significant impact on most organizations in the near future.

The understanding of emerging technologies is also dependent on the viewpoint of the analyst/domain expert \cite{rotolo_what_2015}, and their project objectives. One expert might perceive technology as emerging due to its novelty and anticipated impact on society and the economy, while others might view the same technology as an extension of an already-existing one \cite{rotolo_what_2015}. This lack of consensus over definitions and the variety of perspectives has led to the development of different methodological approaches to detect and analyze emerging technologies. In addition, the absence of comprehensive and accurate historical data on emerging technologies makes the forecasting of these technologies a very challenging task \cite{daim_forecasting_2006}. 

Scientometrics studies, favoring quantitative indicators and employing various data sources such as patents and publications, have been the most common approach for identifying emerging technologies \cite{daim_forecasting_2006}. For example, \cite{porter_technology_1995} applied bibliometrics and proposed a method to monitor and extract insight into emerging technologies from public electronic databases. Lee \cite{lee_how_2008} employed co-word analysis and scientometric indicators to uncover emerging research themes within the realm of information security, analyzing patterns and trends in the field. As another example, in \cite{glanzel_using_2011}, using the notion of \textit{core documents} and cross-citations, authors proposed a bibliometric method for detecting emerging topics. In 2012, \cite{abercrombie_study_2012} conducted an examination of the interconnections among various data sources and proposed a scientometric model aimed at monitoring the emergence of novel technologies. In a more recent study, \cite{wang_tracking_2019} employed bibliometric techniques to investigate emerging technologies in cancer research, focusing on specific types of cancer, i.e., breast and prostate cancers. Due to a lack of consensus in definitions, it is not surprising that these methods for detecting and analyzing emergence often exhibit substantial disparities, even when employing identical or similar techniques \cite{rotolo_what_2015}.

Digital technologies have enhanced the accessibility of massive data, with several science and technology databases available catering to various fields of innovation in which data is continuously increasing in terms of size and variety \cite{bengisu_forecasting_2006}. On the other hand, the existence of advanced computer science algorithms opens up new possibilities for acquiring more profound insights from a diverse set of digital databases \cite{ebadi_detecting_2022}. In \cite{jibu_scientometrics_2018}, authors employed dynamic topic modeling, an unsupervised machine learning technique, to detect and explore the thematic patterns and emerging directions within scientific publications in the field of scientometrics. Ebadi et al. \cite{ebadi_machine_2023} utilized NLP techniques and dynamic topic modeling to analyze the hypersonics research landscape and its temporal evolution. In another study and using patent data \cite{park_technological_2018}, a quantitative forward-looking approach was proposed for identifying technological opportunities that anticipated potential flows of technological knowledge between heterogeneous domains. In a study conducted in 2019 \cite{xu_emerging_2019}, authors built and combined multiple machine learning models and introduced a framework aimed at identifying and anticipating emerging research themes at a thematic level. Using various data sources of diverse types, in \cite{griol-barres_detecting_2020}, a system was introduced that relied on natural language processing (NLP) for identifying weak signals and automatically categorizing the identified keywords. And, using deep learning and weak signal analysis, in \cite{ebadi_detecting_2022}, a multi-layered approach was proposed to detect emerging technologies signals from scientific publications.

In addition to the variety observed in the literature in terms of definitions and methods, conventional approaches employed for identifying emerging technologies have certain constraints. These include relying on simple indicators, necessitating extensive manual involvement, and introducing subjectivity due to heavy reliance on domain expert input at the outset \cite{ebadi_detecting_2022}. Apart from subjectivity, although expert knowledge is a valuable source of information, they may not be aware of emerging technologies outside their specific domain or may overlook innovations they are not familiar with. Moreover, expert insights are often qualitative and challenging to quantify. This can make it difficult to compare and prioritize emerging technologies objectively. Motivated by the recent advances in the field of AI and to address the above-mentioned limitations, in this work, we introduce WISDOM, an automated framework empowered by artificial intelligence, designed to detect emerging research themes through advanced topic modeling and weak signal analysis. In a similar work, Park and Kim \cite{park_study_2021} employed structural topic modeling (STM) \cite{roberts_stm_2019} and weak signal analysis to statistically analyze the temporal shifts in renewable energy topics using publications from 2010 to 2019, and investigated the characteristics of the extracted topics. WISDOM enhances their approach in at least two significant ways: 1) WISDOM utilizes BERTopic \cite{grootendorst_bertopic_2022}, which presents several advantages over STM. BERTopic harnesses pre-trained large language models for language comprehension, enabling it to capture more nuanced semantic relationships and context compared to STM. Furthermore, BERTopic generates representative keywords and document embeddings, resulting in more interpretable topics that facilitate easier understanding and interpretation. And, in addition to depicting shifts in topics over time, the BERTopic model also provides information on time-specific topic representations, enhancing our comprehension of the field's evolution. 2) WISDOM employs an automatic topic labeling engine, which not only enhances the performance and efficiency of the framework but also eliminates the need for manual intervention, thereby automating the pipeline and ensuring objectivity. WISDOM serves as a valuable tool for aiding strategic planners and domain experts in the more efficient recognition and monitoring of trends related to emerging topics while enhancing the effectiveness and objectivity of the detection process. 

\section{Data and Methodology}
\label{sec:data_methods}

\subsection{Data}\label{sec:data}
Researchers communicate their findings, achievements, and scientific breakthroughs to both the scientific community and the broader public through scientific publications \cite{nelkin_performance_1998}, consequently, publications are regarded as the primary output of scientific research \cite{rennie_when_1997}. Hence, as a case technology and to test and validate WISDOM, we focused on scientific publications as the primary data source for the identification of emerging technologies in the field of \textit{underwater sensing}, and we opted for Elsevier's Scopus as our data source. Employing a comprehensive search query carefully crafted by our domain experts\footnote{The search query can be provided upon request.}, we conducted a thorough data collection encompassing all scientific publications on underwater sensing technologies spanning from the year 2004 to 2021.  The dataset we gathered contained all available metadata associated with each of the extracted publications, as provided by the Scopus website. This encompassed, among other details, the paper titles, abstracts, publication dates, list of authors, and their affiliations. Publications lacking either a title or an abstract were excluded from our dataset, and we specifically considered papers authored in the English language. The data collection was executed on September 14, 2022 and the collected dataset contained 9,046 publications.

Data preparation and processing comprised several sequential steps. Initially, we created a new feature named \textit{pubtext} by merging both the title and the abstract for every publication in the dataset. We did not have access to the full text of publications. The abstract encapsulates the entirety of a publication's contribution or content, making it a repository of valuable information. In contrast, a publication's title, despite its brevity, can also offer supplementary informative keywords and/or keyphrases that may not be explicitly found in the abstract section. These considerations drove our decision to combine titles and abstracts, thereby acquiring a more precise representation of the content contained within the publications.

Next, the \textit{pubtext} feature underwent a series of preprocessing steps. First, we converted all text to lowercase, eliminating any case-related inconsistencies. Next, we implemented a comprehensive custom list to remove stop words, enhancing the relevance of the remaining terms. We also addressed special characters, correcting any irregularities that might affect the analysis. Additionally, we removed punctuation marks to streamline the text further. Tokenization was another crucial step, breaking down the text into individual tokens or words for analysis. Lastly, we applied lemmatization, a process that reduces words to their base or dictionary form which results in a more coherent and meaningful representation of the text data. These preprocessing steps collectively contributed to the refinement and standardization of our data for subsequent analysis.

\subsection{Methodology}\label{sec:methods}
\begin{figure}[!ht]
     \centering
     \includegraphics[width=\textwidth]{"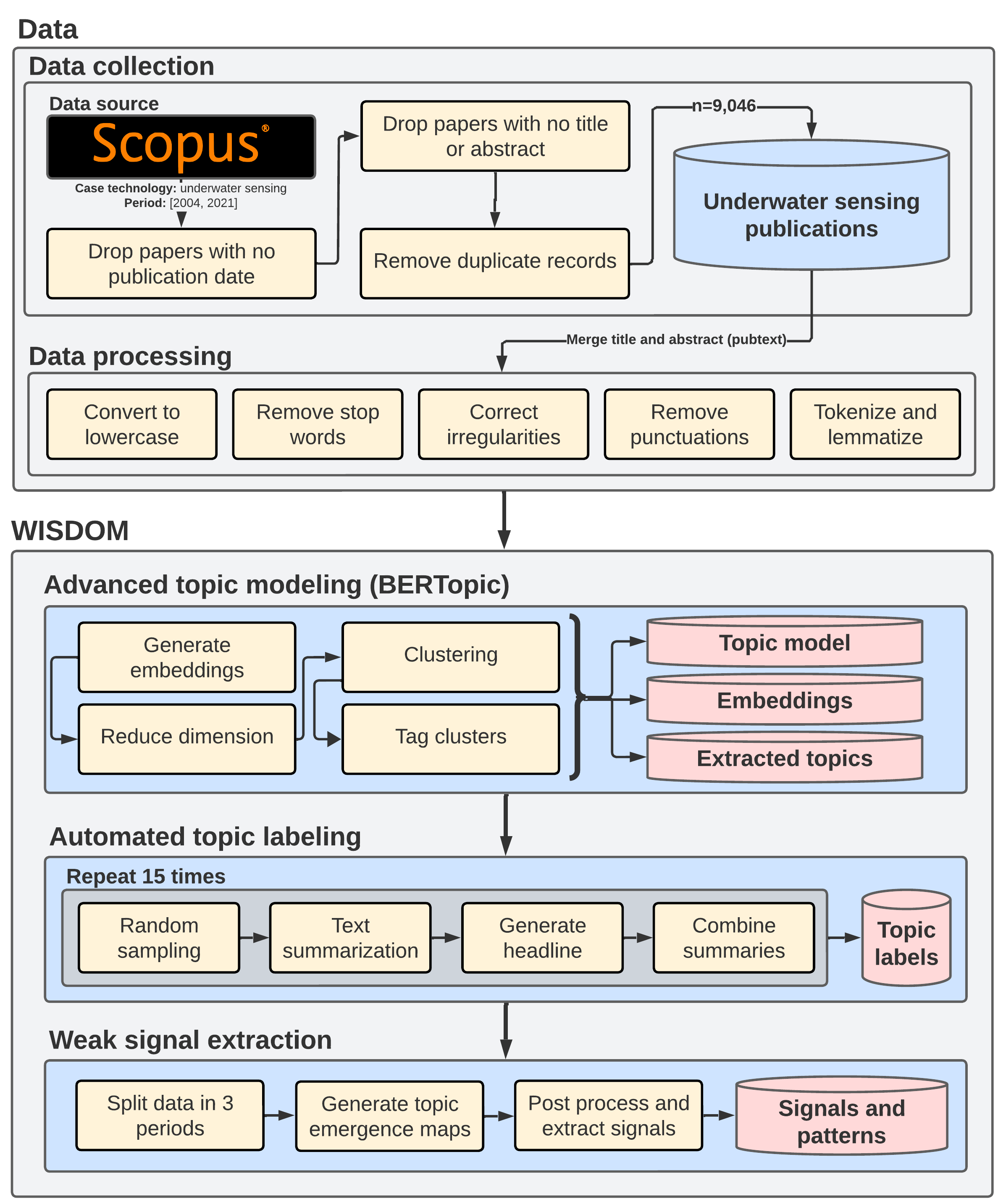"}
    \caption{The high-level conceptual flow of the analytical pipeline.}
    \label{fig1_concep_flow}
\end{figure}

WISDOM's architecture comprises 3 main components: 1) Advanced topic modeling, 2) Automated topic labeling, and 3) Weak signal extraction. This section explains these components in detail. Figure \ref{fig1_concep_flow} illustrates the overarching conceptual flow of the analytical pipeline. The analytical pipeline was developed and executed using the Python programming language. The experiments were performed on an HP laptop equipped with a 2.40 GHz CPU, 16 GB of RAM, and 500 GB of storage capacity. For automated topic labeling, a GPU cluster machine with 4 GPUs and 64 GB of GPU memory was employed.

\subsubsection{Advanced Topic Modeling}\label{sec:bertopic}
WISDOM employs BERTopic \cite{grootendorst_bertopic_2022} to extract the main research themes from the dataset. The primary advantage of the BERTopic model lies in its ability to learn coherent language patterns and excel across a variety of tasks \cite{grootendorst_bertopic_2022}, unlike other common models that tend to specialize in a single area \cite{egger_topic_2022}. Furthermore, following the training of a BERTopic model, it is possible to reduce the number of topics further \cite{sanchez-franco_travelers_2022}, which enables researchers to refine and choose a more precise number of topics based on their objectives. In addition, BERTopic is very flexible as it decouples the document embedding process from topic representation which allows utilizing different preprocessing and fine-tuning techniques.

BERTopic has four major components. Document embeddings are first generated using pre-trained transformer-based language models to obtain document-level information. Leveraging the Sentence-BERT (SBERT) framework \cite{reimers_sentence-bert_2019}, the embedding process generates document representations in a vector space, allowing for semantic comparisons between embeddings. The underlying assumption is that documents sharing the same topic exhibit semantic similarity \cite{grootendorst_bertopic_2022}. Once embeddings are constructed, BERTopic reduces their dimensionality by compressing them into a lower-dimensional space using the uniform manifold approximation and production (UMAP) technique \cite{mcinnes_umap_2020}. UMAP can retain a greater portion of both local and global features of high-dimensional data when projected into lower dimensions. Furthermore, UMAP's flexibility extends to accommodating varying dimensional spaces in different language models due to its lack of computational constraints on embedding dimensions \cite{mcinnes_umap_2020}. Next, HDBSCAN clustering \cite{mcinnes_hdbscan_2017} which is a hierarchical, density-based method is applied to the generated reduced embeddings. HDBSCAN allows the modeling of noise as outliers, ensuring that unrelated documents are not arbitrarily assigned to any cluster, hence; enhancing the quality of topic representations \cite{grootendorst_bertopic_2022}.

Furthermore, reducing high-dimensional embeddings using UMAP can lead to enhanced performance for HDBSCAN, in terms of both clustering accuracy and computational efficiency \cite{allaoui_considerably_2020}. Ultimately topic representations are created through the class-based term frequency-inverse document frequency (c-TF-IDF) procedure. Conventional topic modeling techniques, e.g., latent Dirichlet allocation (LDA) \cite{blei_latent_2003}, are static and cannot capture the sequential organization of documents \cite{ebadi_understanding_2021}. Dynamic topic modeling techniques address this limitation by considering the evolution of topics over time and assessing the degree to which topic representations mirror this evolution. BERTopic achieves this modeling capability by using the c-TF-IDF representations of topics \cite{grootendorst_bertopic_2022}.

In our experiments focusing on underwater sensing publications (as the case technology), following the data cleaning and preprocessing procedures outlined in Section~\ref{sec:data}, we trained the BERTopic model using the lemmatized data. We initially set the number of topics to 100, allowing the model to autonomously determine the optimal number of topics based on the data's characteristics. We also employed an n-gram range of (1,2), which encompassed both single words and bigrams. As a result, the model identified and extracted 73 distinct topics from the corpus which will be presented and discussed in Section~\ref{sec:results}.

\subsubsection{Automated Topic Labeling}\label{sec:autolabel}
Similar to all common topic modeling techniques, BERTopic does not label the extracted topics automatically. In order to enhance the interpretability of the extracted topics and to remove the need for manual intervention, we implemented a multi-layered approach aimed at automatically generating descriptive labels for topics extracted by BERTopic. 

Compressing a topic into a single representative label is a challenging task, often requiring manual input from domain experts. Thanks to recent advancements in the field of natural language processing (NLP), headline generators trained on extensive datasets of news articles are available, capable of producing a concise one-line headline for a provided article. However, these models suffer from one major limitation that is they are trained on sequences with a maximum length of 512 tokens. 

\begin{figure}[!ht]
     \centering
     \begin{subfigure}[b]{0.45\textwidth}
         \centering
         \includegraphics[width=\textwidth]{"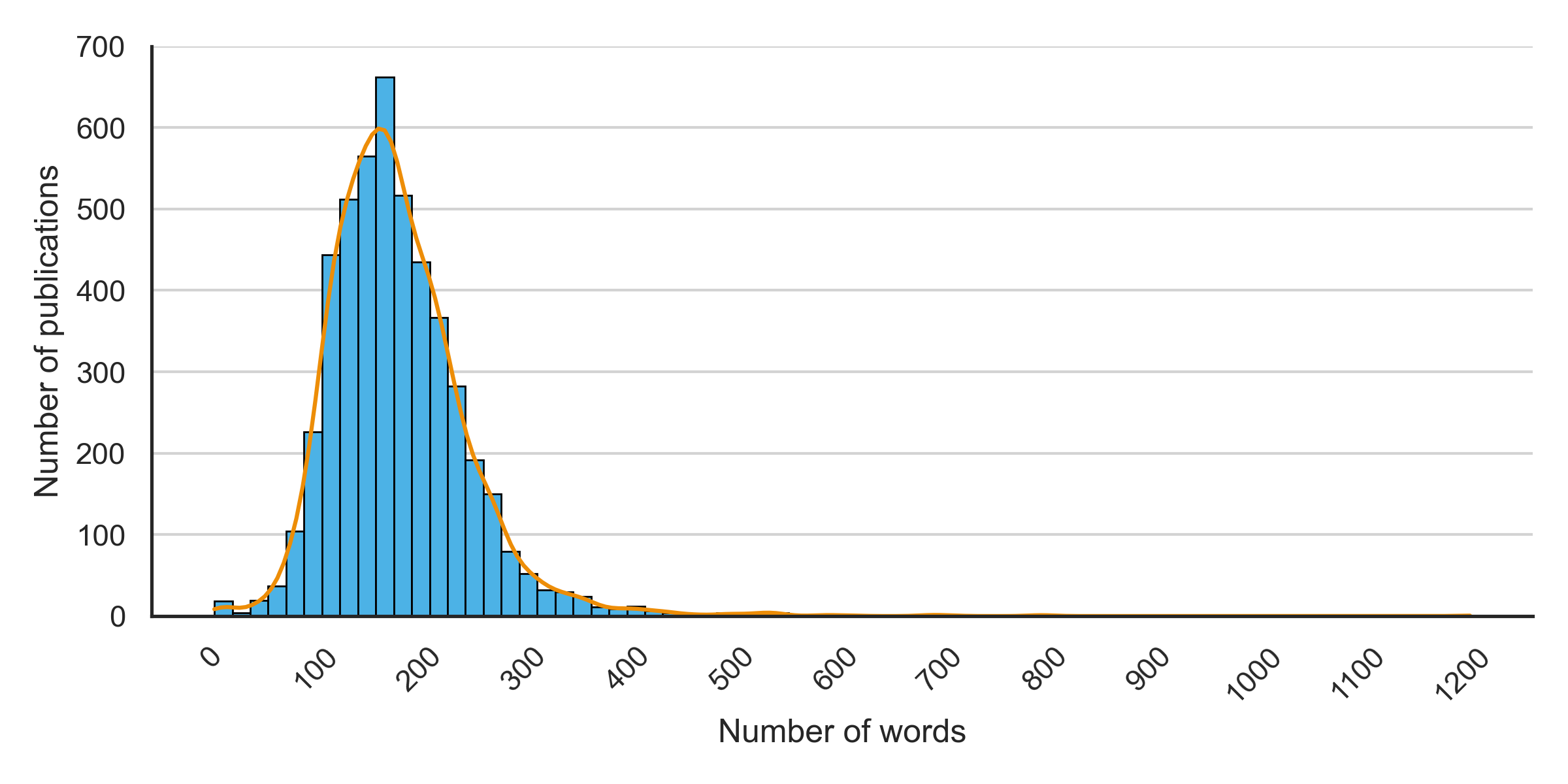"}
         \caption{Number of papers of a given token length}
         \label{fig2a_tokens_papers}
     \end{subfigure}
     \hfill
     \begin{subfigure}[b]{0.45\textwidth}
         \centering
         \includegraphics[width=\textwidth]{"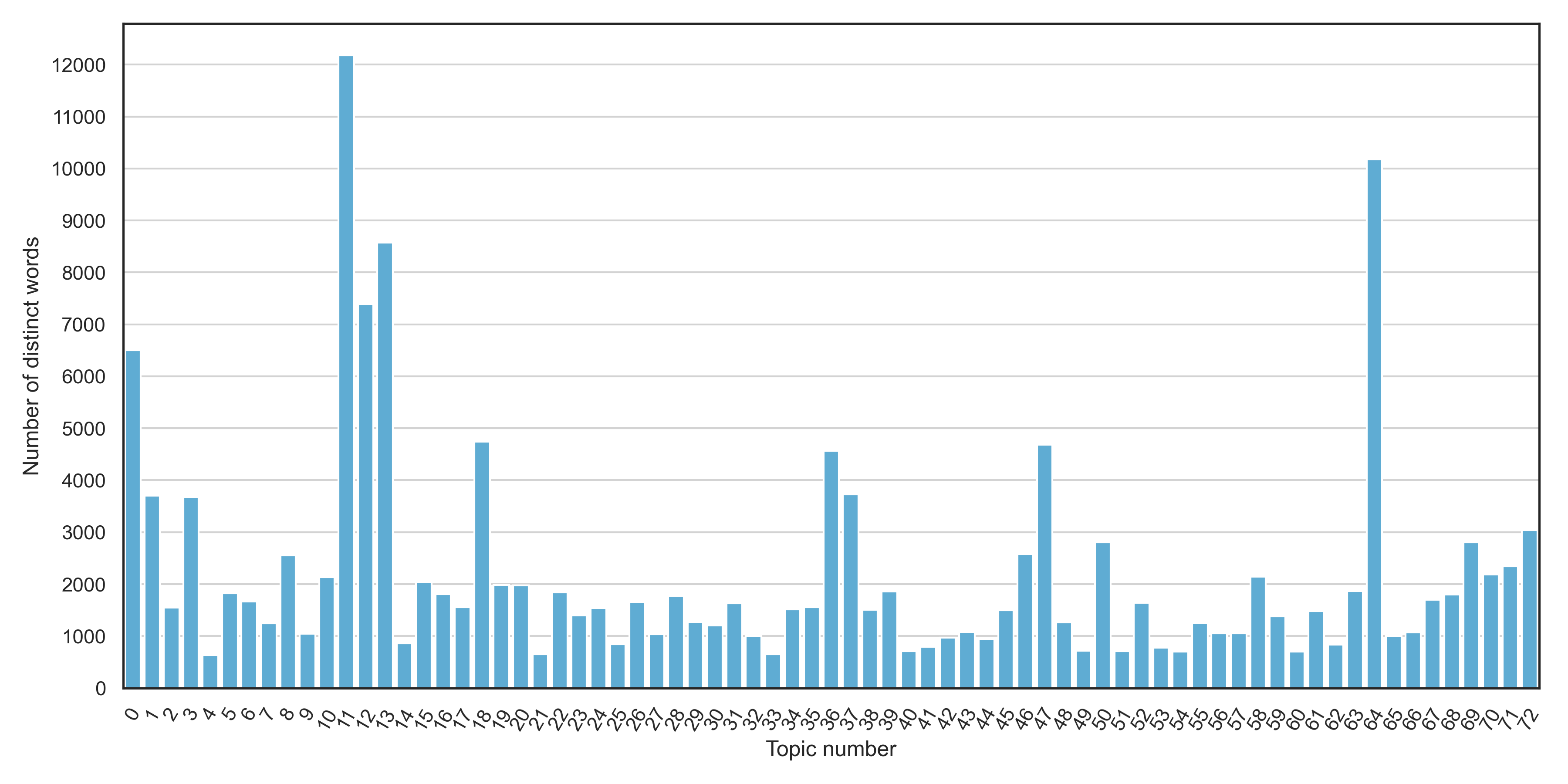"}
         \caption{Number of distinct tokens per topic}
         \label{fig2b_tokens_topics}
     \end{subfigure}
    \caption{Distribution of tokens.}
    \label{fig2_tokens_dist}
\end{figure}

Figure \ref{fig2_tokens_dist} shows the distribution of words in our dataset. As seen in Figure \ref{fig2a_tokens_papers}, papers in the dataset, represented by the \textit{pubtext} feature, are around $100-250$ words in length (median $= 171$), although considerably longer texts are also available in the dataset. On the other hand, a topic contains many papers which makes it impossible to generate a label for topics using headline generators. Figure \ref{fig2b_tokens_topics} shows the number of distinct words per extracted topic from the underwater sensing dataset (median $= 1552$). 

To overcome these limitations and generate a representative label for the extracted topics automatically, we followed an approach that consisted of the following sequential steps. The process is initiated by selecting a random set of papers associated with each of the extracted topics. 

These selected papers were then subjected to summarization using a pre-trained Text-To-Text Transfer Transformer (T5) model \cite{raffel_exploring_2020}. The resulting summaries were combined and using a pre-trained T5-based headline generator model a representative title was generated for the combined summaries. To ensure robustness, the mentioned steps were repeated fifteen times for each topic. Lastly, to guarantee the alignment of the automatically generated labels with the extracted topics, we conducted a review thoroughly assessing both the extracted topics and the automatically generated labels.

\subsubsection{Weak Signal Extraction}\label{sec:weak_signal}
Identifying emerging signals and trends in the ever-evolving technology landscape at an early stage poses a significant challenge. Employing the concept of weak signals within the extracted topics, this study aims to identify research topics that currently exhibit limited attention but hold promise for further development. Ansoff \cite{ansoff_managing_1975} introduced the \textit{weak signal} concept as an alternative or supplement to strategic planning, which, during the 1970s and 1980s, held a dominant role within companies and organizations. At its initial emergence, a weak signal contains vague information, merely hinting at a potential threat or opportunity. Over time, this information gradually expands, delineating the source and attributes of the threat or opportunity, and ultimately, the anticipated outcomes; hence, weak signals can be considered indicators of emerging risks or opportunities in the future \cite{holopainen_weak_2012}. Hiltunen \cite{hiltunen_future_2008} delineated \textit{future signs} as current anomalies and oddities deemed significant in anticipating forthcoming shifts and proposed three dimensions for future signs: (1) \textit{signal}, i.e., quantity and/or visibility of a future sign, (2) \textit{issue}, i.e., events illustrating the dissemination of a future sign, and (3) \textit{interpretation}, i.e. the recipient's comprehension of a future sign's meaning and significance.

In order to quantitatively extract future signals, Yoon proposed a keyword-based text mining methodology, incorporating keyword frequency (as a measure of importance) and document frequencies of keywords (as an indicator of keyword spread) \cite{yoon_detecting_2012}. Although being used in several studies (e.g., \cite{ebadi_detecting_2022,krigsholm_applying_2019,park_analysis_2020}), Yoon's keyword-based approach has two major limitations: (1) signals exclusively present in either the keyword emergence map (KEM) or the keyword issue map (KIM), and not in both of them, are excluded from the future signs. And, (2) as the signals identified as future signs are represented by words, interpreting their meaning could be challenging \cite{park_study_2021}. To address the limitations highlighted, we adopted Park and Kim's approach \cite{park_study_2021} in this study, and employed topic proportions obtained from the BERTopic model instead of relying on word and document frequencies. Signals are derived through the construction of Topic Emergence Maps (TEMs).

\begin{figure}[!ht]
     \centering
     \includegraphics[scale=0.4]{"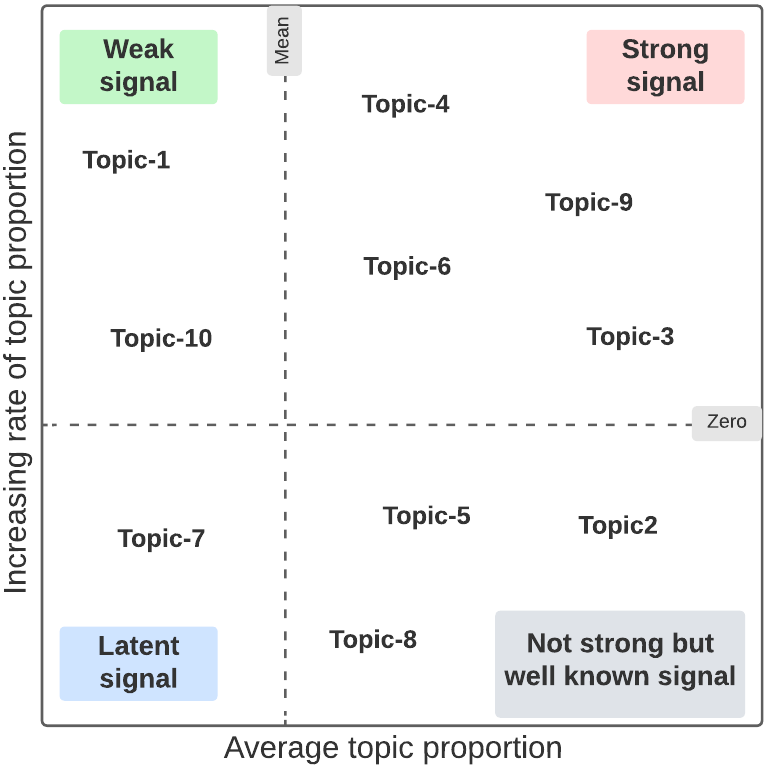"}
    \caption{Topic emergence map (TEM), detecting future signals. The TEM is divided into four quadrants based on the mean of values on the x-axis and a y-axis value of zero.}
    \label{fig3_tem}
\end{figure}

Figure \ref{fig3_tem} illustrates a sample TEM, where the x-axis represents the average topic proportions from the BERTopic model, and the y-axis signifies the rate of increase in topic proportions. The TEM is sectioned into four quadrants as follows, categorized based on the x-axis mean and a y-axis value of zero: (1) The top-right region designates \textit{strong} signals, encompassing research topics with both high average proportion and growth rate. (2) The top-left region signifies \textit{weak} signals, involving topics with a low average proportion yet high growth rate. (3) Terms located in the bottom-left zone are classified as \textit{latent} signals. And, (4) The bottom-right area characterizes signals identified as \textit{not strong but well-known (NSWK)}. For instance, in Figure \ref{fig3_tem}, Topic-10 is classified as weak, Topic-3 as strong, Topic-7 as latent, and Topic-8 as an NSWK signal. 

As described in Section \ref{sec:data}, publications related to underwater sensing technologies from 2004 to 2021 were collected for this study. To track the temporal evolution and changes in patterns of signals, we segmented the examined period into three six-year intervals, i.e., $[2004,2009]$, $[2010,2015]$, and $[2016,2021]$. We will refer to these intervals as $p_1$, $p_2$, and $p_3$, respectively, in the rest of the text. TEMs were then built for each period and a signal label was assigned to each topic within the TEM across the examined periods, namely, $p_1$ to $p_3$.

\section{Results}
\label{sec:results}

\subsection{Topics}\label{subsec:topics}
As outlined in Section \ref{sec:bertopic}, the BERTopic model \cite{grootendorst_bertopic_2022} was utilized to extract research themes from the publications on underwater sensing. BERTopic assigns -1 to all outlier topics, hence; we removed it. Table \ref{tab1:topics} presents the remaining topics along with their document counts. The first column represents the topic number, followed by its label and respective publication count. Topic65, identified as \quotes{submarine image enhancement and color correction} has the highest document count and three topics, i.e., Topic34, Topic55, and Topic63, have the lowest.

\begin{longtblr}[
  caption = {Extracted topics and their respective publication count.},
  label = {tab1:topics},
  note{} = { \footnotesize * The first column (\#) indicates topic number. \\ \footnotesize ** For list of abbreviations, please see \ref{sec:appendix_a}. }
]{
  colspec = {|l|l|l|},
  rowhead = 1,
  hlines,
} 
\textbf{\#} & \textbf{Topic Label} & \textbf{Ct.}\\
 1 & Arctic Ocean, Ice, and Oil & 146 \\
 2 & WSNs, Data and Applications & 96 \\
 3 & Canadian Navy & 24 \\
 4 & Underwater Fish Tracking & 82 \\
 5 & Design and Manufacturing of New Aircraft with Enhanced Capacity & 14 \\
 6 & Maritime Security, Big Data and Sensors & 18 \\
 7 & UASNs & 39 \\
 8 & Australia's Anti-Submarine Warfare Program & 23 \\
 9 & Efficient Underwater Acoustics Networking & 91 \\
 10 & MAC Protocol for USNs & 26 \\
 11 & Laser Radar, Ultrasonics & 32 \\
 12 & Maritime Remote Sensing and Surveillance & 638 \\
 13 & Maritime Aircraft & 285 \\
 14 & Underwater 3D Acoustic Imaging System & 393 \\
 15 & Oceanography  & 14 \\
 16 & China's Military Modernization & 26 \\
 17 & Submarine Warfare, History & 25 \\
 18 & UWSNs, Ocean Exploration & 24 \\
 19 & Underwater Robotics & 160 \\
 20 & Marine Mammal Detection and Tracking & 30 \\
 21 & Underwater Imaging & 47 \\
 22 & Detection of Explosive and Other Hazardous Substances & 14 \\
 23 & Detecting Coral Reefs & 36 \\
 24 & Detecting and Tracking of Underwater Vehicles/Objects & 22 \\
 25 & Submarine Image Enhancement & 32 \\
 26 & Mobile Anchor in USNs & 14 \\
 27 & Sea Floor Characterization & 27 \\
 28 & Submarine Object Detection, Detecting Small Objects & 13 \\
 29 & Deep Sea Robotics & 31 \\
 30 & Sonar Image, Processing Techniques and Quality Evaluation & 21 \\
 31 & Deep Sea Neutron Telescope & 22 \\
 32 & Submarine Co2 Monitoring & 26 \\
 33 & Coap Congestion Control Scheme for UWSNs & 20 \\
 34 & Energy Efficiency in UASNs & 12 \\
 35 & Restoration and Enhancement of Underwater Image & 28 \\
 36 & Routing Protocols for UWSNs & 28 \\
 37 & Research on USNs & 176 \\
 38 & Underwater Laser Serial Imaging Sensors & 95 \\
 39 & Underwater Acoustic Communication Networks & 25 \\
 40 & Submarine Volcano & 21 \\
 41 & Low-Power UWSNs & 14 \\
 42 & SDRT & 13 \\
 43 & Geocast Technique for USNs & 16 \\
 44 & Subsea Infrastructure Monitoring & 16 \\
 45 & Sub-Regional Queries in UWSNs & 18 \\
 46 & Wireless Sensor Charging & 41 \\
 47 & Submarine Unexploded Ordnance & 35 \\
 48 & Submarine Oil Monitoring & 140 \\
 49 & Geospatial Routing in USNs & 31 \\
 50 & HHVBF Routing & 14 \\
 51 & Sonar Monitoring and Tracking & 66 \\
 52 & Submarine Ghost Imaging & 13 \\
 53 & Clustering in USNs & 43 \\
 54 & PULRP for USNs & 14 \\
 55 & Umimo-MAC, New Medium Access Control Protocol & 12 \\
 56 & Low-Power Communication & 30 \\
 57 & Deep Learning for UOD & 19 \\
 58 & Submarine Laser Based Remote Imaging and Tracking & 19 \\
 59 & Subsea Cable & 33 \\
 60 & CTDA, Localization of USNs & 25 \\
 61 & Searching Submarines, Submarine Magnetic Field & 14 \\
 62 & DCS for UASNs & 27 \\
 63 & Mini-Automated Underwater Vehicle & 12 \\
 64 & LiDAR Signal Processing for UOD & 45 \\
 65 & Submarine Image Enhancement and Color Correction & 839 \\
 66 & Sensor Fault Detection & 22 \\
 67 & Underwater Target Tracking and Simulation & 22 \\
 68 & Submarine Mine Detection & 34 \\
 69 & Sonar Imaging for Underwater Targets & 42 \\
 70 & Methods to Route WSNs & 78 \\
 71 & Comparative Studies on UWSNs & 46 \\
 72 & Energy-Efficient Algorithms for UWSNs & 49 \\
 73 & Localization Methods for UWSNs & 81 \\
\end{longtblr}

\begin{figure}[!ht]
     \centering
     \includegraphics[scale=0.17]{"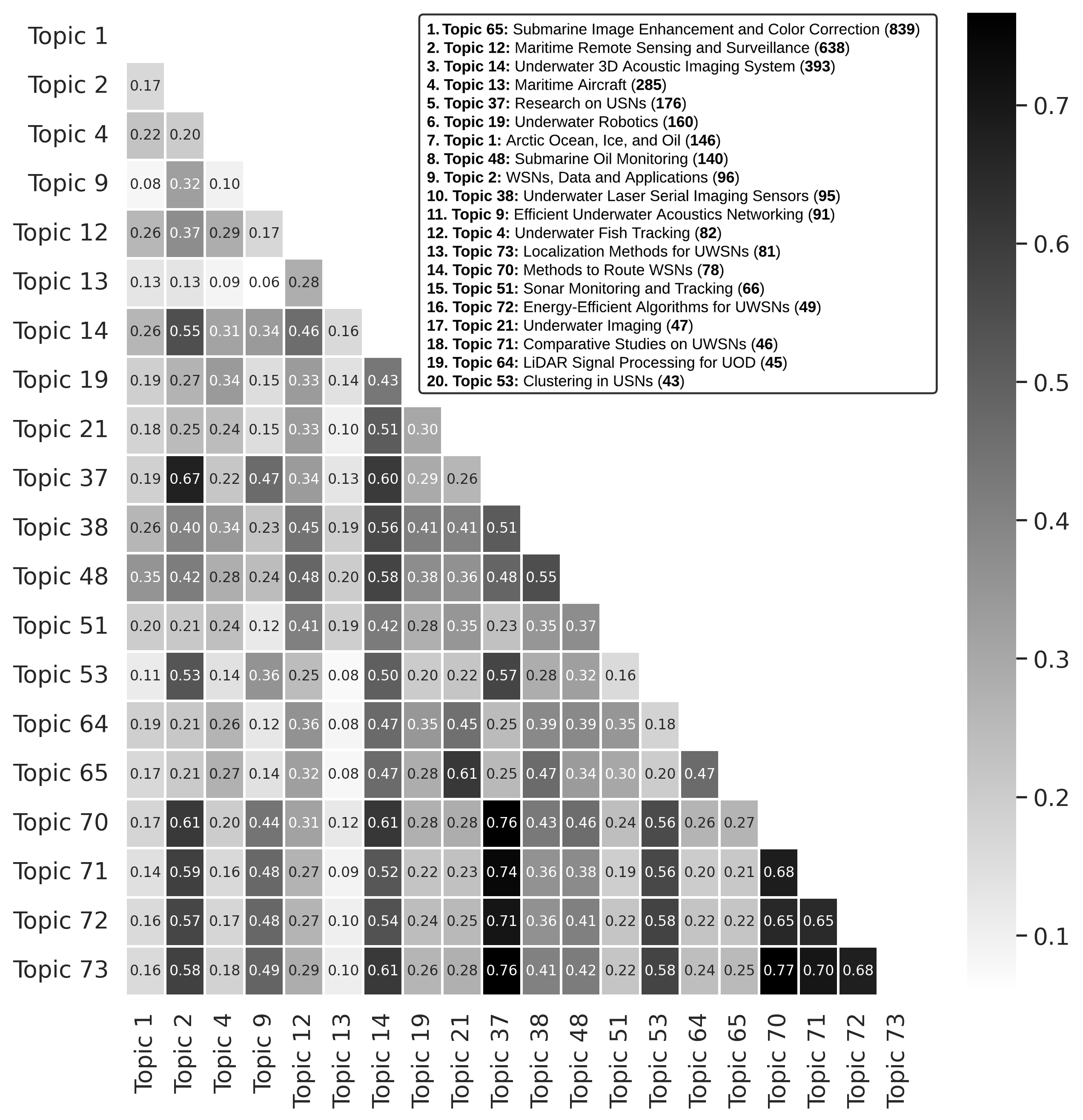"}
    \caption{Similarity matrix, top-20 most frequent topics. The numbers in parentheses within the legend indicate the document counts corresponding to each topic. The values in the cells denote the similarity score.}
    \label{fig4_topic_sim}
\end{figure}

Figure \ref{fig4_topic_sim} illustrates the heatmap depicting the similarity among the top 20 topics with the highest document counts. The heatmap was generated using a cosine similarity matrix computed from the topic embeddings. As observed, the greatest similarity is noted between Topic70 (\quotes{methods to route WSNs}) and Topic73 (\quotes{localization methods for UWSNs}), and the lowest similarity is seen between Topic9 (\quotes{efficient underwater acoustics networking}) and Topic13 (\quotes{maritime aircraft}).

\subsection{Signals}\label{subsec:signs}
\begin{figure}[!ht]
     \centering
     \includegraphics[scale=0.15]{"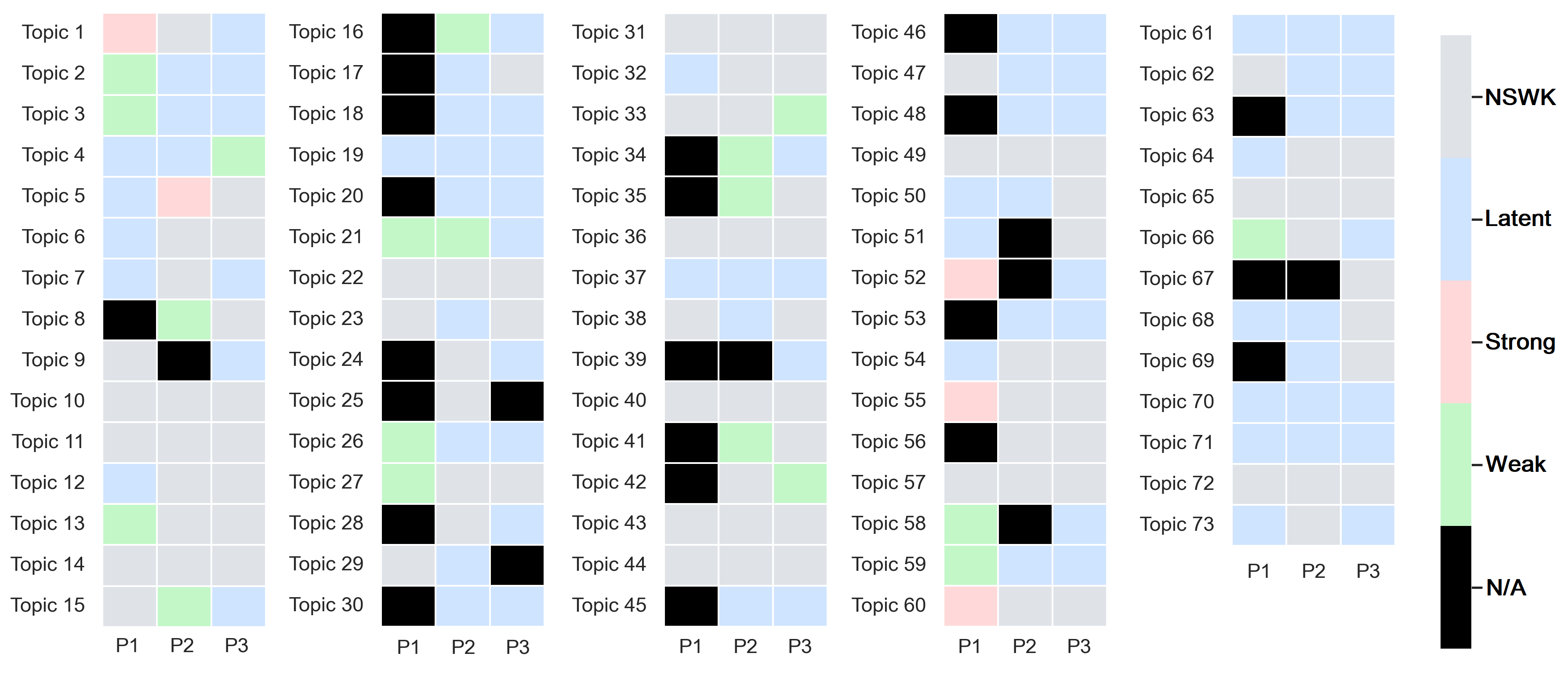"}
    \caption{Topics' signals in $P_1$, $P_2$, and $P_3$. Black cells indicate no signal.}
    \label{fig5_sig_trend}
\end{figure}

Figure \ref{fig5_sig_trend} shows the signal patterns for the extracted research topics in the three six-year intervals, i.e., $p_1=[2004,2009]$, $p_2=[2010,2015]$, and $p_3=[2016,2021]$, as described in Section \ref{sec:weak_signal}. As seen, none of the weak signal topics (highlighted in green in the figure) transitioned into strong signals (highlighted in red) during the subsequent periods. Topic25, labeled as \quotes{submarine image enhancement}, solely appeared in the period $P_2$, whereas Topic39, identified as \quotes{underwater acoustic communication networks}, and Topic67, denoted as \quotes{underwater target tracking and simulation}, exclusively emerged in the period $P_3$. Eighteen topics exhibited a consistent pattern throughout the entire analyzed period. Among them, five topics—Topics19, 37, 61, 70, and 71—were identified as latent signals from $P_1$ to $P_3$. The remaining thirteen topics, i.e., Topic10, 11, 14, 22, 31, 36, 40, 43, 44, 49, 57, 65, and 72, were classified as NSWK signals throughout the entire period. We will discuss the weak and strong signals in the following sections.

\subsubsection{Weak Signals}\label{subsubsec:weak}
Eighteen topics—Topics2, 3, 4, 8, 13, 15, 16, 21, 26, 27, 33, 34, 35, 41, 42, 58, 59, and 66—appeared as weak signals in at least one of the three examined periods. Among them, only Topic21 (\quotes{underwater imaging}) emerged as a weak signal in two consecutive periods, namely $P_1$ and $P_2$. Four topics—Topics2, 4, 13, and 21—out of these eighteen topics were among the most frequent topics (refer to Figure \ref{fig4_topic_sim}).

\begin{figure}[hbt!]
     \centering
     \begin{subfigure}[b]{0.3\textwidth}
         \centering
         \includegraphics[width=\textwidth]{"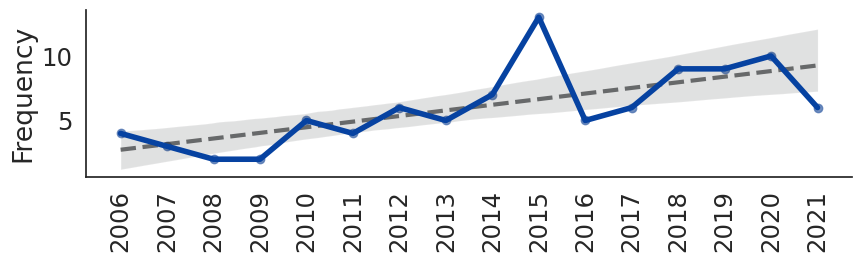"}
         \caption{Topic 2 ($P_1$)}
         \label{fig6:topic2}
     \end{subfigure}
     \hfill
     \begin{subfigure}[b]{0.3\textwidth}
         \centering
         \includegraphics[width=\textwidth]{"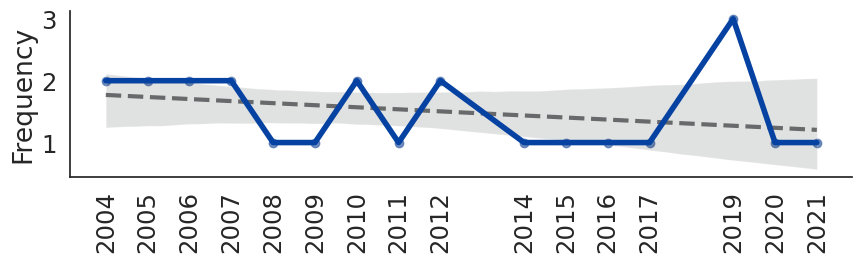"}
         \caption{Topic 3 ($P_1$)}
         \label{fig6:topic3}
     \end{subfigure}
     \hfill
     \begin{subfigure}[b]{0.3\textwidth}
         \centering
         \includegraphics[width=\textwidth]{"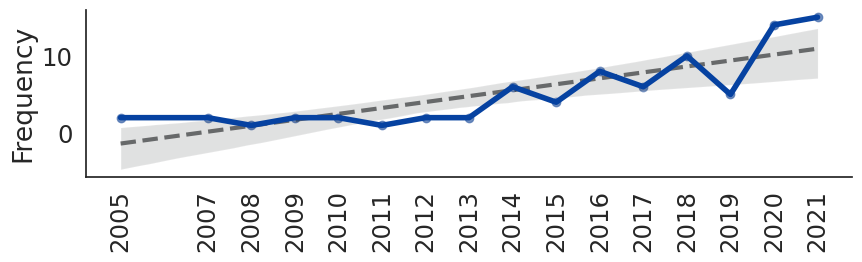"}
         \caption{Topic 4 ($P_3$)}
         \label{fig6:topic4}
     \end{subfigure}
     
     \bigskip
     \begin{subfigure}[b]{0.3\textwidth}
         \centering
         \includegraphics[width=\textwidth]{"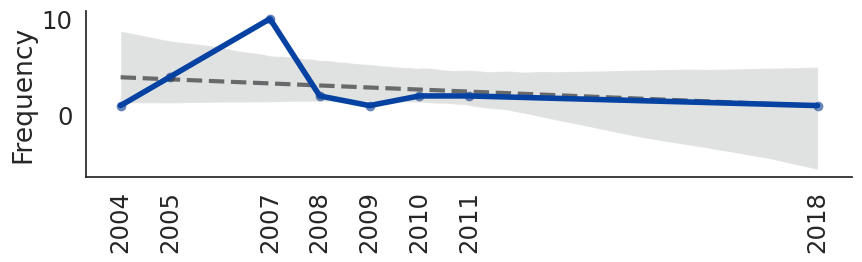"}
         \caption{Topic 8 ($P_2$)}
         \label{fig6:topic8}
     \end{subfigure}
     \hfill
     \begin{subfigure}[b]{0.3\textwidth}
         \centering
         \includegraphics[width=\textwidth]{"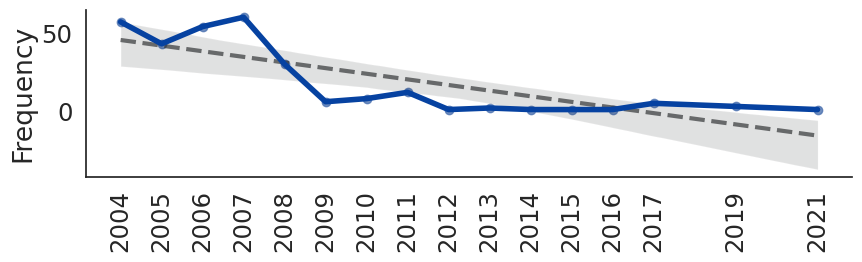"}
         \caption{Topic 13 ($P_1$)}
         \label{fig6:topic13}
     \end{subfigure}
     \hfill
     \begin{subfigure}[b]{0.3\textwidth}
         \centering
         \includegraphics[width=\textwidth]{"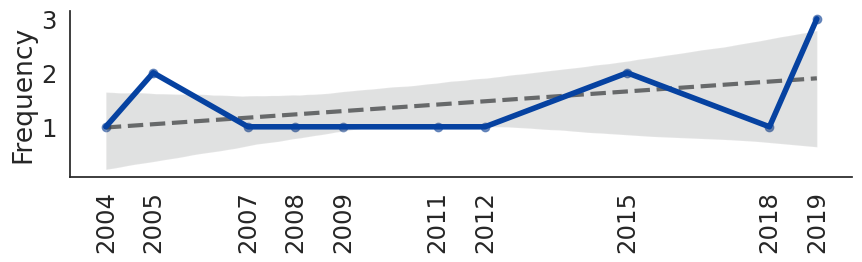"}
         \caption{Topic 15 ($P_2$)}
         \label{fig6:topic15}
     \end{subfigure}
     
     \bigskip
     \begin{subfigure}[b]{0.3\textwidth}
         \centering
         \includegraphics[width=\textwidth]{"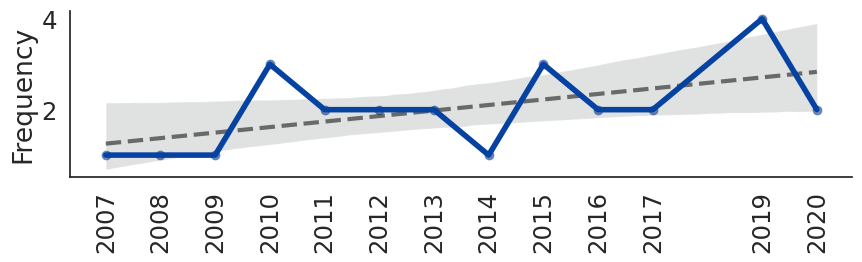"}
         \caption{Topic 16 ($P_2$)}
         \label{fig6:topic16}
     \end{subfigure}
     \hfill
     \begin{subfigure}[b]{0.3\textwidth}
         \centering
         \includegraphics[width=\textwidth]{"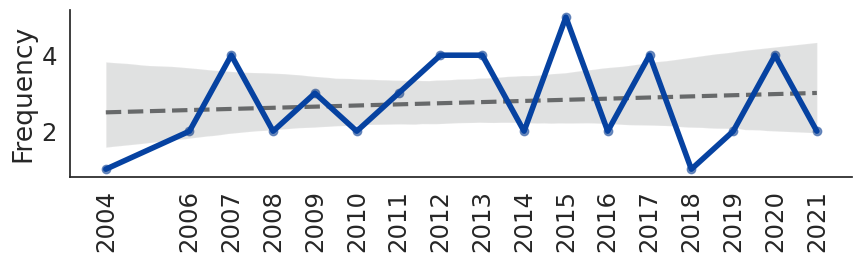"}
         \caption{Topic 21 ($P_1,P2$)}
         \label{fig6:topic21}
     \end{subfigure}
     \hfill
     \begin{subfigure}[b]{0.3\textwidth}
         \centering
         \includegraphics[width=\textwidth]{"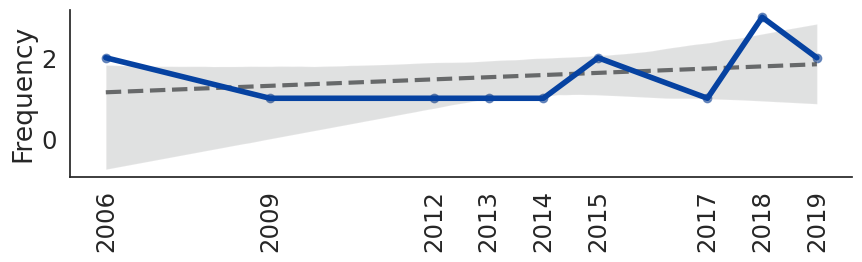"}
         \caption{Topic 26 ($P_1$)}
         \label{fig6:topic26}
     \end{subfigure}

     \bigskip
     \begin{subfigure}[b]{0.3\textwidth}
         \centering
         \includegraphics[width=\textwidth]{"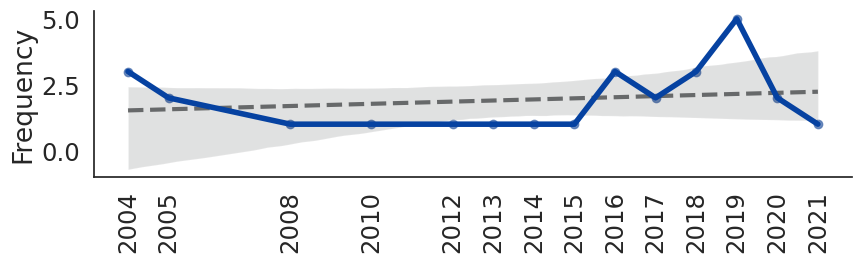"}
         \caption{Topic 27 ($P_1$)}
         \label{fig6:topic27}
     \end{subfigure}
     \hfill
     \begin{subfigure}[b]{0.3\textwidth}
         \centering
         \includegraphics[width=\textwidth]{"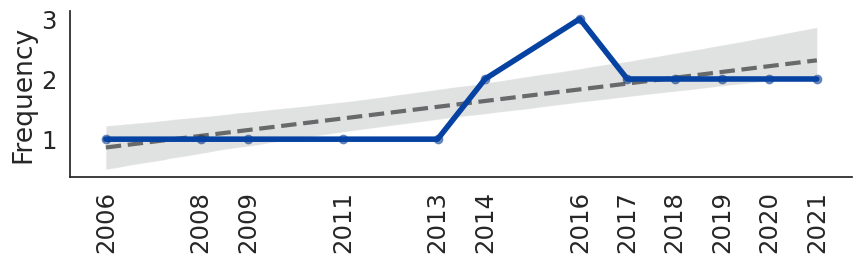"}
         \caption{Topic 33 ($P_3$)}
         \label{fig6:topic33}
     \end{subfigure}
     \hfill
     \begin{subfigure}[b]{0.3\textwidth}
         \centering
         \includegraphics[width=\textwidth]{"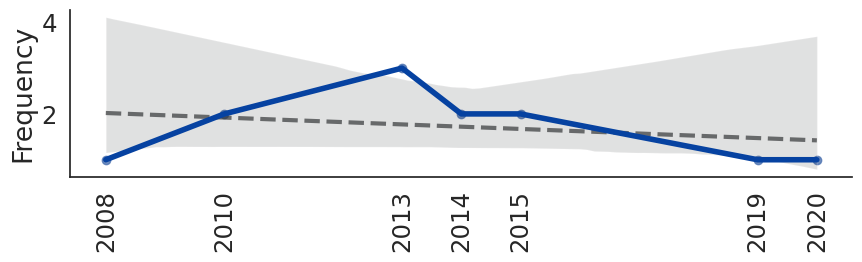"}
         \caption{Topic 34 ($P_2$)}
         \label{fig6:topic34}
     \end{subfigure}

     \bigskip
     \begin{subfigure}[b]{0.3\textwidth}
         \centering
         \includegraphics[width=\textwidth]{"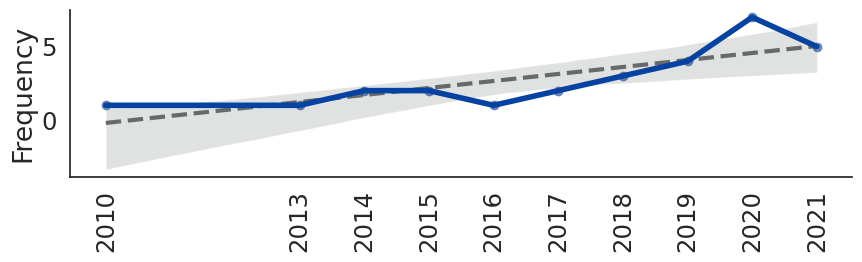"}
         \caption{Topic 35 ($P_2$)}
         \label{fig6:topic35}
     \end{subfigure}
     \hfill
     \begin{subfigure}[b]{0.3\textwidth}
         \centering
         \includegraphics[width=\textwidth]{"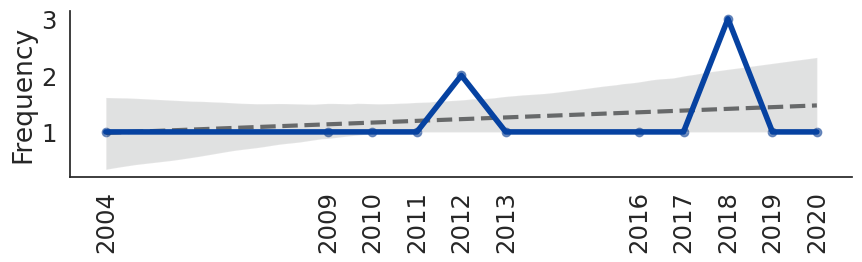"}
         \caption{Topic 41 ($P_2$)}
         \label{fig6:topic41}
     \end{subfigure}
     \hfill
     \begin{subfigure}[b]{0.3\textwidth}
         \centering
         \includegraphics[width=\textwidth]{"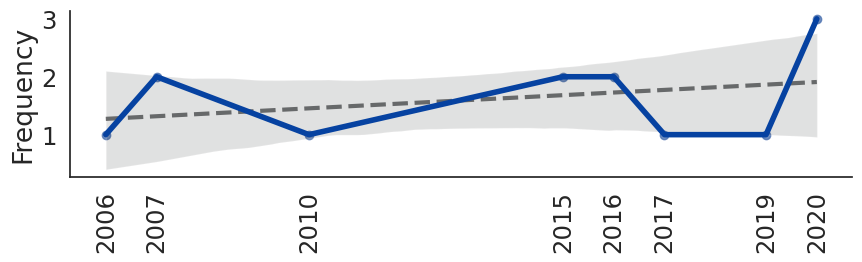"}
         \caption{Topic 42 ($P_3$)}
         \label{fig6:topic42}
     \end{subfigure}

     \bigskip
     \begin{subfigure}[b]{0.3\textwidth}
         \centering
         \includegraphics[width=\textwidth]{"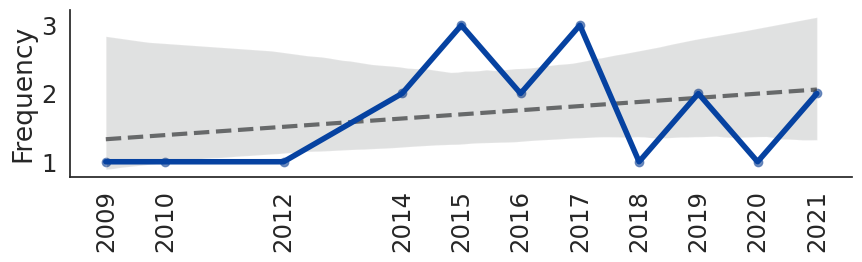"}
         \caption{Topic 58 ($P_1$)}
         \label{fig6:topic58}
     \end{subfigure}
     \hfill
     \begin{subfigure}[b]{0.3\textwidth}
         \centering
         \includegraphics[width=\textwidth]{"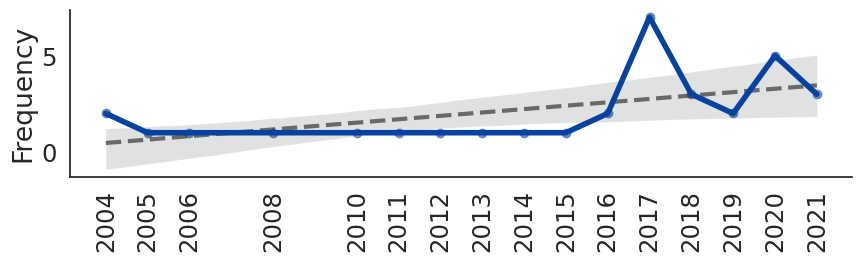"}
         \caption{Topic 59 ($P_1$)}
         \label{fig6:topic59}
     \end{subfigure}
     \hfill
     \begin{subfigure}[b]{0.3\textwidth}
         \centering
         \includegraphics[width=\textwidth]{"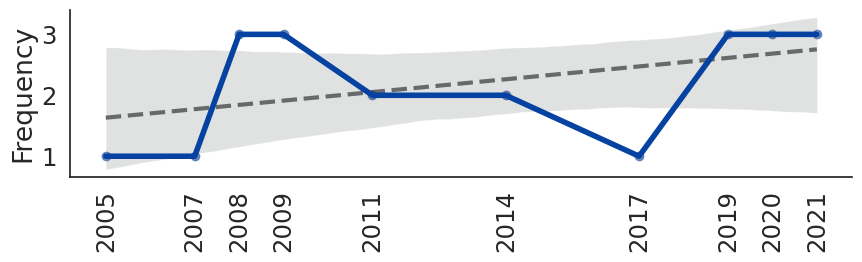"}
         \caption{Topic 66 ($P_1$)}
         \label{fig6:topic66}
     \end{subfigure}
    
    \caption{The temporal evolution of research topics identified as weak signals across various periods (solid blue line). The corresponding periods are indicated in parentheses alongside the topics. The gray dashed line represents the trend line, with the shaded area highlighting the 95\% confidence interval.}
    \label{fig6:weak_evolution}
\end{figure}

Figure \ref{fig6:weak_evolution} depicts the temporal progression of the weak signal research topics (solid blue line), accompanied by a dashed gray trend line encircled by a shaded area denoting the 95\% confidence interval. The x-axis displays only the years corresponding to generated data points. It is important to observe the interval variances when interpreting the sub-plots, as topics may not commence and/or conclude at identical times. As seen, Topics2, 4, 15, 16, 26, 27, 33, 35, 42, 58, 59 and 66 have experienced an upward trend in their most recent intervals whereas topics3, 13, and 34 have declined. Certain topics, such as Topics8 and 13, encountered few peaks before and/or after maintaining a constant trend. Topics8, 21, and 41 exhibited a nearly constant trend throughout the analyzed period.

Furthermore, besides illustrating changes in topics over time, the BERTopic model also furnishes details regarding time-specific topic representations. Table~\ref{tab:weak_keywords} presents the top three representative keywords specific to time intervals corresponding to the top three most recent peaks for each topic. The expanded forms of the acronyms in the table are listed in \ref{sec:appendix_a}. In addition to the keywords and peaks, the intervals between the peaks can also be informative, potentially demonstrating the time needed for a shift within a research topic over time. For instance, Topic21 (underwater imaging) exhibited fluctuations throughout the entire period, with relatively minor gaps between peaks and lows, while Topic41 (low-power UWSNs) experienced only two peaks separated by a six-year interval (see Figure \ref{fig6:weak_evolution} and Table \ref{tab:weak_keywords}), potentially suggesting varying levels of activity within their respective fields.

\begin{table}[hbt!]
\begin{tabular}{|l|l|l|}
\hline
\textbf{\#} & \textbf{\{{Keywords\}}$_{year}$} \\ 
\hline
 2 & \small \{wireless, aggregation, sensor\}$_{2015}$, \{wireless, satcom, network\}$_{2018}$, \\ 
   & \{iomt, iot, wireless\}$_{2020}$ \\ 
 3 & \small \{cuwps, canadian, berthed\}$_{2010}$, \{container, canadian, icargo\}$_{2012}$, \\ 
   & \{canadian, voyage, politics\}$_{2019}$ \\ 
 4 & \small \{fish, edna, aquaculture\}$_{2018}$, \{fish, aquaculture, video\}$_{2020}$, \\ 
   & \{fish, specie, aquaculture\}$_{2021}$ \\ 
 8 & \small \{australia, helicopter, dmo\}$_{2007}$ \\ 
 13 & \small \{navy, aircraft, warfare\}$_{2004}$, \{navy, warfare, aircraft\}$_{2006}$, \\ 
   & \{aircraft, navy, antisubmarine\}$_{2007}$ \\ 
 15 & \small \{worldwar, revolution, ultrasonic\}$_{2005}$, \{equation, sonar, ocean\}$_{2015}$, \\ 
   & \{anthropocene, munk, scapa\}$_{2019}$ \\ 
 16 & \small \{china, di, military\}$_{2010}$, \{chick, parent, young\}$_{2015}$, \\ 
   & \{chinese, military, foreign\}$_{2019}$ \\ 
 21 & \small \{waveform, mimo, wideband\}$_{2015}$, \{apit, nonlocal, image\}$_{2017}$, \\ 
   & \{artemide, pure, spectral\}$_{2020}$ \\ 
 26 & \small \{cooperative, spacetime, dfe\}$_{2006}$, \{ambient, ofdm, collaborative\}$_{2015}$, \\ & \{bistatic, target, probability\}$_{2018}$ \\ 
 27 & \small \{sav, aquatic, vegetation\}$_{2004}$, \{label, speed, motor\}$_{2016}$, \\ 
   & \{shoal, filter, bearing\}$_{2019}$ \\
 33 & \small \{cns, node, greedy\}$_{2016}$ \\ 
 34 & \small \{esvbf, head, cluster\}$_{2013}$ \\ 
 35 & \small \{image, segmentation, multilevel\}$_{2020}$ \\ 
 41 & \small \{modem, itaca, power\}$_{2012}$, \{dualhop, lowpower, transceiver\}$_{2018}$ \\ 
 42 & \small \{revelation, dtn, uwasn\}$_{2015}$, \{auv, resurfacing, inter\}$_{2016}$, \\ 
   & \{captain, uoasns, node\}$_{2020}$ \\ 
 58 & \small \{singlephoton, ultrasonic, transducer\}$_{2015}$, \{module, hotwire, buoyancy\}$_{2017}$ \\ 
 59 & \small \{sand, scour, bridge\}$_{2017}$, \{scour, bolted, beach\}$_{2020}$ \\ 
 66 & \small \{fault, multitarget, channel\}$_{2019}$, \{fault, kirchhoff, forwardscattering\}$_{2020}$, \\ & \{keel, uict, fault\}$_{2021}$ \\ 
\hline
\end{tabular}
\vspace{0.2cm}
\caption{Weak topic signals' keywords representatives of their recent three peaks.}
\label{tab:weak_keywords}
\end{table}

\subsubsection{Strong Signals}\label{subsubsec:strong}
\begin{figure}[hbt!]
     \centering
     \begin{subfigure}[b]{0.3\textwidth}
         \centering
         \includegraphics[width=\textwidth]{"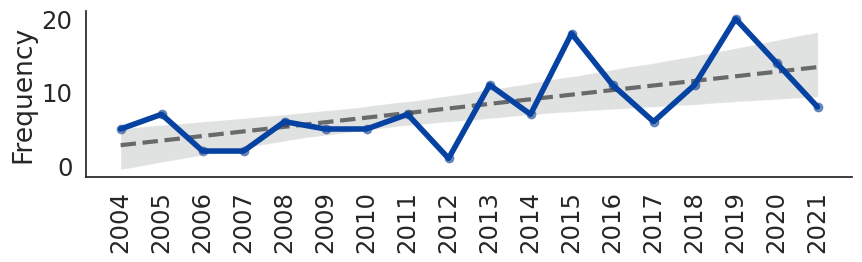"}
         \caption{Topic 1 ($P_1$)}
         \label{fig7:topic1}
     \end{subfigure}
     \hfill
     \begin{subfigure}[b]{0.3\textwidth}
         \centering
         \includegraphics[width=\textwidth]{"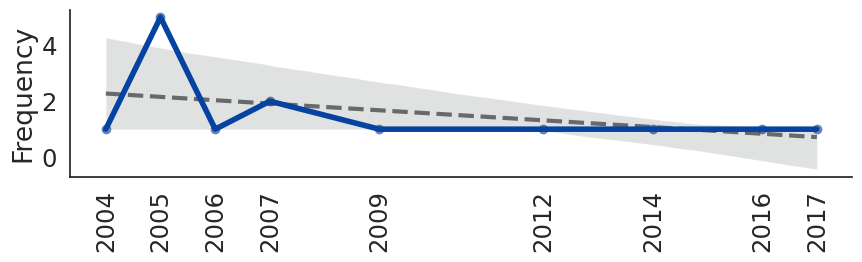"}
         \caption{Topic 5 ($P_2$)}
         \label{fig7:topic5}
     \end{subfigure}
     \hfill
     \begin{subfigure}[b]{0.3\textwidth}
         \centering
         \includegraphics[width=\textwidth]{"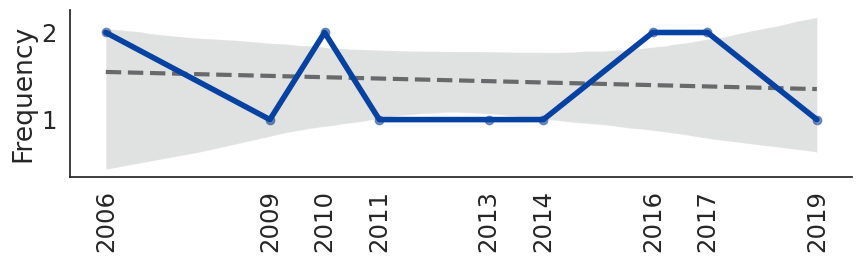"}
         \caption{Topic 52 ($P_1$)}
         \label{fig7:topic52}
     \end{subfigure}
     
     \bigskip
     \begin{subfigure}[b]{0.3\textwidth}
         \centering
         \includegraphics[width=\textwidth]{"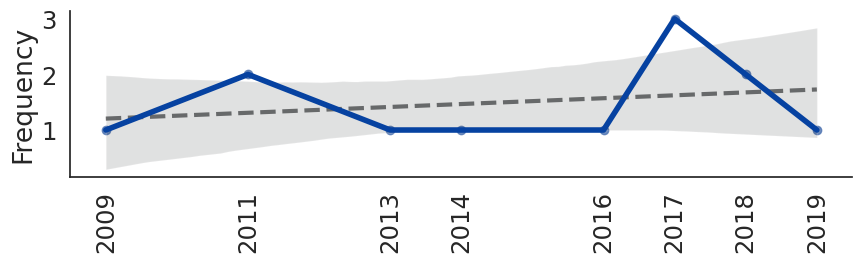"}
         \caption{Topic 55 ($P_1$)}
         \label{fig7:topic55}
     \end{subfigure}
     \hfill
     \begin{subfigure}[b]{0.3\textwidth}
         \centering
         \includegraphics[width=\textwidth]{"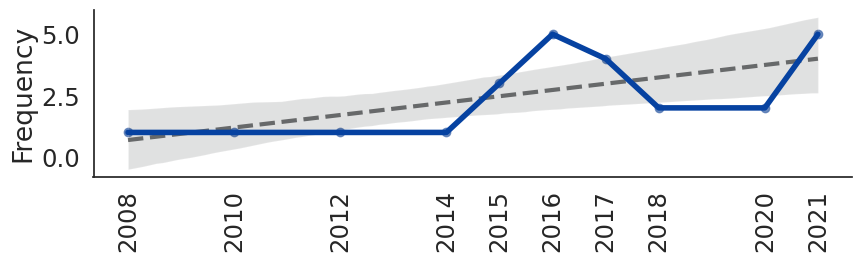"}
         \caption{Topic 60 ($P_1$)}
         \label{fig7:topic60}
     \end{subfigure}

    \caption{The temporal evolution of research topics identified as strong signals across various periods (solid blue line). The corresponding periods are indicated in parentheses alongside the topics. The gray dashed line represents the trend line, with the shaded area highlighting the 95\% confidence interval.}
    \label{fig7:strong_evolution}
\end{figure}

Five topics—Topics1, 5, 52, 55, and 60—emerged as strong signals in at least one of the periods, with all appearing in $P_1$ except for Topic5, which appeared in $P_2$. Of these strong signal topics, only Topic1 (\quotes{design and manufacturing of new aircraft with enhanced capacity}) is among the top-20 most frequent topics (see Figure \ref{fig4_topic_sim}). Figure \ref{fig7:strong_evolution} shows the temporal evolution of strong signal research topics. Observing the trends, three topics—Topics1, 55, and 60—demonstrated an overall upward trajectory, while the remaining two topics—Topics5 and 52—exhibited a decline. Notably, Topic60 is projected to experience growth in the foreseeable future, while others are anticipated to maintain a steady or decreasing trend. Among these five topics, Topic1 has exhibited the most fluctuations and the least constant intervals compared to the other four topics, suggesting its enduring significance over time.

Table \ref{tab:strong_keywords} presents the top three time-specific keywords corresponding to the top three most recent peaks for each of the strong signal topics. The full expansions of the acronyms in the table are listed in \ref{sec:appendix_a}. It is noteworthy that Topic1, exhibiting the most fluctuations compared to others (Figure \ref{fig7:strong_evolution}), showcases the most consistent set of representative keywords across its peaks, suggesting a greater concentration within the corresponding research field.

\begin{table}[hbt!]
\begin{tabular}{|l|l|l|}
\hline
\textbf{\#} & \textbf{\{{Keywords\}}$_{year}$} \\ 
\hline
 1 & \small \{ice, oil, snow\}$_{2013}$, \{ice, arctic, oil\}$_{2015}$, \{ice, oil, arctic\}$_{2019}$ \\ 
 5 & \small \{mission, upgrade, helicopter\}$_{2005}$ \\ 
 52 & \small \{spatial, frequencywave, array\}$_{2010}$, \{narrowband, tracker, neutron\}$_{2016}$, \\    & \{ghost, function, optical\}$_{2017}$ \\ 
 55 & \small \{unimomac, access, function\}$_{2011}$, \{aliveinrange, queue, access\}$_{2017}$ \\ 
 60 & \small \{asn, estimation, localization\}$_{2016}$, \{locatization, optimizer, dfo\}$_{2021}$ \\ 
\hline
\end{tabular}
\vspace{0.2cm}
\caption{Strong topic signals' keywords representatives of their recent three peaks.}
\label{tab:strong_keywords}
\end{table}

\section{Discussion and Conclusion}
\label{sec:discussion}
In this work, we presented WISDOM, an automated framework empowered by artificial intelligence, devised to detect emerging research themes through the integration of multiple techniques such as advanced topic modeling and weak signal analysis. As a case study, we evaluate WISDOM's efficacy in identifying emerging research themes and trends in underwater sensing technologies using scientific literature spanning from 2004 to 2021.

Eighteen research topics were identified as weak signals in at least one of the three periods, namely $P_1$, $P_2$, and $P_3$. As depicted in Figure \ref{fig6:weak_evolution}, among the eighteen weak signal topics, only Topics4 (underwater fish tracking), 33 (CoAP congestion control scheme for UWSNs), and 42 (SDRT) emerged in $P_3$, with all of them anticipated to sustain growth in the following years. In recent years, underwater fish tracking has captured the interest of researchers due to its potential to revolutionize our understanding of aquatic ecosystems and fisheries management \cite{saleh_applications_2024}. Advanced sensing technologies, such as acoustic telemetry and satellite tagging \cite{matley_global_2022}, and digital technologies, such as deep learning and computer vision \cite{saleh_applications_2024}, have enabled researchers to track fish movements in real-time over large spatial scales, providing valuable insights into migration patterns, habitat utilization, and population dynamics. 

Congestion management in wireless sensor networks (Topic33) is also a challenging and important problem that recently garnered significant attention from researchers \cite{pandey_exploratory_2020}. Advanced sensing technologies and the proliferation of interconnected underwater devices have heightened the need for efficient congestion control mechanisms to ensure reliable and timely data transmission in underwater environments. This scheme offers promising solutions by leveraging the lightweight and energy-efficient characteristics of the Constrained Application Protocol (CoAP) \cite{makarem_design_2022}, tailored specifically for the unique constraints and conditions of UWSNs, thereby improving network performance and facilitating a wide range of underwater applications, from the Internet of Underwater Things (IoUT) \cite{seo_new_2020} to the Internet of Robotic Things (IoRT) \cite{kabir_internet_2023}. Reliable data transmission in underwater environments is crucial. With the advent of advanced sensing technologies and the increasing deployment of underwater sensor networks, there is a growing demand for robust communication protocols capable of ensuring data integrity and reliability despite harsh underwater conditions. Segmented Data Reliable Transfer (SDRT, Topic42) \cite{xie_sdrt_2010} offers a promising solution by employing segmented data transmission techniques tailored to the unique characteristics of underwater communication, thereby enhancing the efficiency and reliability of data transfer in underwater research applications.

As depicted in Figure \ref{fig5_sig_trend}, five topics—Topics1, 5, 52, 55, and 60—emerged as strong signals, with all appearing in the $P_1$ period except for Topic5, which emerged in $P_2$. From Figure \ref{fig7:strong_evolution}, only Topics1 (arctic ocean, ice, oil) and 60 (CTDA, localization of USNs) experienced an overall increasing trend, with the growing trend expected to continue for Topic60. The Arctic Ocean has been an intriguing research topic for researchers in underwater sensing technologies due to several factors such as environmental concerns \cite{piermattei_cost-effective_2018}, resource exploration \cite{cross_innovative_2015}, geopolitical significance \cite{huebert_arctic_2022}, and technological challenges. Scientists have aimed to address several critical research challenges such as ice monitoring and prediction\cite{chi_prediction_2017}, oil spill detection and mapping \cite{mooradian_spectral_2014}, and underwater mapping and exploration using advanced sonar systems and AUVs \cite{kunz_deep_2008}, and contribute to the sustainable management of Arctic resources and ecosystems through innovative underwater sensing technologies. Precise localization and tracking of USNs (Topic60) could also enhance environmental monitoring, resource exploration, and maritime security. Recent research in this area focuses on developing advanced acoustic and electromagnetic localization techniques \cite{su_review_2020}, improving underwater communication systems \cite{yan_localization_2021}, and integrating AUVs for efficient data collection and network maintenance \cite{yan_auv-aided_2020}. 

The BERTopic model \cite{grootendorst_bertopic_2022} provides time-specific keyword sets for each extracted topic (see Tables \ref{tab:weak_keywords} and \ref{tab:strong_keywords}) that enable experts to gain insights into the evolving nature of topics over time. By analyzing the keywords associated with each topic at different time intervals, we can track changes in the focus or emphasis of discussions within a particular topic area. This can help identify emerging trends, shifting research interests, or evolving terminology within a field. As an example, Topic1 (refer to Table \ref{tab:strong_keywords}) represents a tightly focused research area, whereas Topic4 exemplifies a transition in data sources (Table \ref{tab:weak_keywords}), and Topic2 illustrates shifts in technology (refer to Table \ref{tab:weak_keywords}). Comparing keyword sets across different time periods can reveal patterns or correlations between topics, providing a deeper understanding of the dynamic nature of research discourse and facilitating more informed decision-making in various domains. Additionally, the temporal evolution curves (see Figures \ref{fig6:weak_evolution} and \ref{fig7:strong_evolution}), characterized by peaks and lows with intervals between consecutive peaks, offer valuable insights into the dynamics of topic popularity or relevance over time. By examining these patterns, experts can discern trends, identify periods of heightened interest or activity, and uncover shifts in research focus and the time needed for a shift within a given topic area. 

While manual identification of emerging research topics can provide valuable insights, it is often limited by its time-consuming nature, subjectivity, and potential inefficiencies in handling large volumes of data. In addition, manual identification may lack consistency and standardization across different domain experts, leading to variations in the identification process and potentially impacting the reliability of the results. Automated approaches, such as using machine learning and natural language processing techniques, can help mitigate these disadvantages by providing more scalable, objective, and efficient methods for identifying emerging technologies. WISDOM offers strategic planners and domain experts a streamlined approach to identifying and monitoring emerging research topics by rapidly processing extensive datasets, unveiling latent cross-disciplinary patterns, and delivering impartial insights, thereby enhancing the efficacy and objectivity of trend detection. Lastly, while showcased in this paper within the domain of underwater sensing technologies, WISDOM is applicable to other domains as well.

\section{Limitations and Future Work}
\label{sec:lim}
In this work, we tested the performance of our proposed approach on scientific publications related to the \textit{underwater sensing} technologies. Although publications are one of the main sources for reporting scientific discoveries, not all emerging technologies are immediately documented in scientific publications. Testing the proposed approach on other data sources such as industry reports, funding proposals, or other academic/non-academic sources, could be a future research direction. Some innovative projects might be kept confidential or proprietary. The proposed approach cannot capture these hidden innovations. The definition of \textit{emerging technology} can vary, and different researchers may have different criteria for what constitutes emergence. Not all published research papers are of equal quality or significance. Focusing only on high-impact research and excluding lower-quality or less influential works could be a future direction. Our data only covered English papers and research published in other languages was not considered. Future research may apply the proposed approach to research published in other languages. Finally, future research endeavors could involve monitoring the identified weak signal topics and investigating their dynamics and patterns as they evolve into strong signals.

\appendix

\section{Abbreviations}
\label{sec:appendix_a}
\begin{itemize}
    \setlength{\itemsep}{1pt}
    \setlength{\parskip}{0pt}
    \setlength{\parsep}{0pt}
    \item Acoustic Sensor Network (\textbf{ASN})
    \item Autonomous Underwater Vehicle (\textbf{AUV})
    \item Constrained Node Set (\textbf{CNS})
    \item Constrained Application Protocol (\textbf{CoAP}) 
    \item Connected Tree Depth Adjustment (\textbf{CTDA})
    \item The Canadian Underwater Protection System (\textbf{CUwPS})
    \item Data-Centric Storage (\textbf{DCS})
    \item Decision Feedback Equalization (\textbf{DFE})
    \item Derivative-Free Optimization (\textbf{DFO})
    \item The Defence Materiel Organisation (\textbf{DMO})
    \item Delay and Disruption Tolerant Network (\textbf{DTN})
    \item Environmental DeoxyriboNucleic Acid (\textbf{eDNA})
    \item Hop-by-Hop Vector-Based Forwarding (\textbf{HHVBF})
    \item Internet of Multimedia Things (\textbf{IoMT})
    \item Internet of Things (\textbf{IoT})
    \item Internet of Robotic Things (\textbf{IoRT})
    \item Internet of Underwater Things (\textbf{IoUT})
    \item Media Access Control (\textbf{MAC})
    \item Multiple-Input Multiple-Output (\textbf{MIMO})
    \item Orthogonal Frequency Division Multiplexing (\textbf{OFDM})
    \item Path Unconscious Layered Routing Protocol (\textbf{PULRP})
    \item Submerged Aquatic Vegetation (\textbf{SAV})
    \item Segmented Data Reliable Transfer (\textbf{SDRT})
    \item Underwater Acoustic Sensor Network (\textbf{UASN})
    \item Underwater Information and Communication Technology (\textbf{UICT})
    \item Underwater Optical-Acoustic Sensor Network (\textbf{UOASN})
    \item Underwater Object Detection (\textbf{UOD})
    \item Underwater Sensor Network (\textbf{USN})
    \item Underwater Acoustic Sensor Network (\textbf{UWASN})
    \item Underwater Wireless Sensor Network (\textbf{UWSN})
    \item Wireless Sensor Network (\textbf{WSN})
\end{itemize}

\bibliographystyle{unsrt}  
\bibliography{main}  

\begin{thebibliography}{10}

\bibitem{kenski_tecnologias_2013}
Vani~Moreira Kenski.
\newblock {\em Tecnologias e ensino presencial e a distância}.
\newblock Papirus Editora, May 2013.
\newblock Google-Books-ID: WHeADwAAQBAJ.

\bibitem{viet_analyzing_2021}
Nguyen~Thanh Viet and Vladislav Gneushev.
\newblock Analyzing and {Forecasting} {Emerging} {Technology} {Trends} by {Mining} {Web} {News}.
\newblock In Alla~G. Kravets, Maxim Shcherbakov, Danila Parygin, and Peter~P. Groumpos, editors, {\em Creativity in {Intelligent} {Technologies} and {Data} {Science}}, Communications in {Computer} and {Information} {Science}, pages 55--69, Cham, 2021. Springer International Publishing.

\bibitem{muhlroth_systematic_2018}
Christian Mühlroth and Michael Grottke.
\newblock A systematic literature review of mining weak signals and trends for corporate foresight.
\newblock {\em Journal of Business Economics}, 88(5):643--687, July 2018.

\bibitem{noh_identifying_2016}
Heeyong Noh, Young-Keun Song, and Sungjoo Lee.
\newblock Identifying emerging core technologies for the future: {Case} study of patents published by leading telecommunication organizations.
\newblock {\em Telecommunications Policy}, 40(10):956--970, October 2016.

\bibitem{dotsika_identifying_2017}
Fefie Dotsika and Andrew Watkins.
\newblock Identifying potentially disruptive trends by means of keyword network analysis.
\newblock {\em Technological Forecasting and Social Change}, 119:114--127, June 2017.

\bibitem{ebadi_detecting_2022}
Ashkan Ebadi, Alain Auger, and Yvan Gauthier.
\newblock Detecting emerging technologies and their evolution using deep learning and weak signal analysis.
\newblock {\em Journal of Informetrics}, 16(4):101344, November 2022.

\bibitem{zamani_developing_2022}
Mehdi Zamani, Haydar Yalcin, Ali~Bonyadi Naeini, Gordana Zeba, and Tugrul~U Daim.
\newblock Developing metrics for emerging technologies: identification and assessment.
\newblock {\em Technological Forecasting and Social Change}, 176:121456, March 2022.

\bibitem{grootendorst_bertopic_2022}
Maarten Grootendorst.
\newblock {BERTopic}: {Neural} topic modeling with a class-based {TF}-{IDF} procedure, March 2022.
\newblock arXiv:2203.05794 [cs].

\bibitem{rotolo_what_2015}
Daniele Rotolo, Diana Hicks, and Ben~R. Martin.
\newblock What is an emerging technology?
\newblock {\em Research Policy}, 44(10):1827--1843, December 2015.

\bibitem{porter_measuring_2002}
Alan~L Porter, J~David Roessner, Xiao-Yin Jin, and Nils~C Newman.
\newblock Measuring national ‘emerging technology’ capabilities.
\newblock {\em Science and Public Policy}, 29(3):189--200, June 2002.

\bibitem{boon_exploring_2008}
Wouter Boon and Ellen Moors.
\newblock Exploring emerging technologies using metaphors – {A} study of orphan drugs and pharmacogenomics.
\newblock {\em Social Science \& Medicine}, 66(9):1915--1927, May 2008.

\bibitem{small_identifying_2014}
Henry Small, Kevin~W. Boyack, and Richard Klavans.
\newblock Identifying emerging topics in science and technology.
\newblock {\em Research Policy}, 43(8):1450--1467, October 2014.

\bibitem{veletsianos_emergence_2016}
George Veletsianos.
\newblock {\em Emergence and {Innovation} in {Digital} {Learning}: {Foundations} and {Applications}}.
\newblock Athabasca University Press, June 2016.
\newblock Google-Books-ID: BD7ADAAAQBAJ.

\bibitem{daim_forecasting_2006}
Tugrul~U. Daim, Guillermo Rueda, Hilary Martin, and Pisek Gerdsri.
\newblock Forecasting emerging technologies: {Use} of bibliometrics and patent analysis.
\newblock {\em Technological Forecasting and Social Change}, 73(8):981--1012, October 2006.

\bibitem{porter_technology_1995}
Alan~L. Porter and Michael~J. Detampel.
\newblock Technology opportunities analysis.
\newblock {\em Technological Forecasting and Social Change}, 49(3):237--255, July 1995.

\bibitem{lee_how_2008}
Woo Lee.
\newblock How to identify emerging research fields using scientometrics: {An} example in the field of {Information} {Security}.
\newblock {\em Scientometrics}, 76(3):503--525, July 2008.
\newblock Publisher: Akadémiai Kiadó, co-published with Springer Science+Business Media B.V., Formerly Kluwer Academic Publishers B.V. Section: Scientometrics.

\bibitem{glanzel_using_2011}
Wolfgang Glänzel and Bart Thijs.
\newblock Using ‘core documents’ for detecting and labelling new emerging topics.
\newblock {\em Scientometrics}, 91(2):399--416, December 2011.
\newblock Publisher: Akadémiai Kiadó, co-published with Springer Science+Business Media B.V., Formerly Kluwer Academic Publishers B.V. Section: Scientometrics.

\bibitem{abercrombie_study_2012}
Robert~K. Abercrombie, Akaninyene~W. Udoeyop, and Bob~G. Schlicher.
\newblock A study of scientometric methods to identify emerging technologies via modeling of milestones.
\newblock {\em Scientometrics}, 91(2):327--342, January 2012.
\newblock Publisher: Akadémiai Kiadó, co-published with Springer Science+Business Media B.V., Formerly Kluwer Academic Publishers B.V. Section: Scientometrics.

\bibitem{wang_tracking_2019}
Xiaoyu Wang, Jingjing Guo, Dongxiao Gu, Ying Yang, Xuejie Yang, and Keyu Zhu.
\newblock Tracking knowledge evolution, hotspots and future directions of emerging technologies in cancers research: a bibliometrics review.
\newblock {\em Journal of Cancer}, 10(12):2643--2653, June 2019.

\bibitem{bengisu_forecasting_2006}
Murat Bengisu and Ramzi Nekhili.
\newblock Forecasting emerging technologies with the aid of science and technology databases.
\newblock {\em Technological Forecasting and Social Change}, 73(7):835--844, September 2006.

\bibitem{jibu_scientometrics_2018}
Mari Jibu and Yoshiyuki Osabe.
\newblock {\em Scientometrics}.
\newblock BoD – Books on Demand, July 2018.
\newblock Google-Books-ID: fy6RDwAAQBAJ.

\bibitem{ebadi_machine_2023}
Ashkan Ebadi, Alain Auger, and Yvan Gauthier.
\newblock Machine {Learning} {Insights} into {Hypersonics} {Research} {Evolution}: {A} 21st {Century} {Perspective}.
\newblock {\em Archives of Advanced Engineering Science}, pages 1--16, October 2023.

\bibitem{park_technological_2018}
Inchae Park and Byungun Yoon.
\newblock Technological opportunity discovery for technological convergence based on the prediction of technology knowledge flow in a citation network.
\newblock {\em Journal of Informetrics}, 12(4):1199--1222, November 2018.

\bibitem{xu_emerging_2019}
Shuo Xu, Liyuan Hao, Xin An, Guancan Yang, and Feifei Wang.
\newblock Emerging research topics detection with multiple machine learning models.
\newblock {\em Journal of Informetrics}, 13(4):100983, November 2019.

\bibitem{griol-barres_detecting_2020}
Israel Griol-Barres, Sergio Milla, Antonio Cebrián, Huaan Fan, and Jose Millet.
\newblock Detecting {Weak} {Signals} of the {Future}: {A} {System} {Implementation} {Based} on {Text} {Mining} and {Natural} {Language} {Processing}.
\newblock {\em Sustainability}, 12(19):7848, January 2020.
\newblock Number: 19 Publisher: Multidisciplinary Digital Publishing Institute.

\bibitem{park_study_2021}
Chankook Park and Minkyu Kim.
\newblock A {Study} on the {Characteristics} of {Academic} {Topics} {Related} to {Renewable} {Energy} {Using} the {Structural} {Topic} {Modeling} and the {Weak} {Signal} {Concept}.
\newblock {\em Energies}, 14(5):1497, January 2021.
\newblock Number: 5 Publisher: Multidisciplinary Digital Publishing Institute.

\bibitem{roberts_stm_2019}
Margaret~E. Roberts, Brandon~M. Stewart, and Dustin Tingley.
\newblock stm: {An} {R} {Package} for {Structural} {Topic} {Models}.
\newblock {\em Journal of Statistical Software}, 91:1--40, October 2019.

\bibitem{nelkin_performance_1998}
Dorothy Nelkin.
\newblock The performance of science.
\newblock {\em The Lancet}, 352(9131):893--894, September 1998.
\newblock Publisher: Elsevier.

\bibitem{rennie_when_1997}
Drummond Rennie, Veronica Yank, and Linda Emanuel.
\newblock When {Authorship} {Fails}: {A} {Proposal} to {Make} {Contributors} {Accountable}.
\newblock {\em JAMA}, 278(7):579--585, August 1997.

\bibitem{egger_topic_2022}
Roman Egger and Joanne Yu.
\newblock A {Topic} {Modeling} {Comparison} {Between} {LDA}, {NMF}, {Top2Vec}, and {BERTopic} to {Demystify} {Twitter} {Posts}.
\newblock {\em Frontiers in Sociology}, 7, 2022.

\bibitem{sanchez-franco_travelers_2022}
Manuel~J. Sánchez-Franco and Manuel Rey-Moreno.
\newblock Do travelers' reviews depend on the destination? {An} analysis in coastal and urban peer-to-peer lodgings.
\newblock {\em Psychology \& Marketing}, 39(2):441--459, 2022.
\newblock \_eprint: https://onlinelibrary.wiley.com/doi/pdf/10.1002/mar.21608.

\bibitem{reimers_sentence-bert_2019}
Nils Reimers and Iryna Gurevych.
\newblock Sentence-{BERT}: {Sentence} {Embeddings} using {Siamese} {BERT}-{Networks}, August 2019.
\newblock arXiv:1908.10084 [cs].

\bibitem{mcinnes_umap_2020}
Leland McInnes, John Healy, and James Melville.
\newblock {UMAP}: {Uniform} {Manifold} {Approximation} and {Projection} for {Dimension} {Reduction}, September 2020.
\newblock arXiv:1802.03426 [cs, stat].

\bibitem{mcinnes_hdbscan_2017}
Leland McInnes, John Healy, and Steve Astels.
\newblock hdbscan: {Hierarchical} density based clustering.
\newblock {\em The Journal of Open Source Software}, 2(11):205, March 2017.

\bibitem{allaoui_considerably_2020}
Mebarka Allaoui, Mohammed~Lamine Kherfi, and Abdelhakim Cheriet.
\newblock Considerably {Improving} {Clustering} {Algorithms} {Using} {UMAP} {Dimensionality} {Reduction} {Technique}: {A} {Comparative} {Study}.
\newblock In Abderrahim El~Moataz, Driss Mammass, Alamin Mansouri, and Fathallah Nouboud, editors, {\em Image and {Signal} {Processing}}, Lecture {Notes} in {Computer} {Science}, pages 317--325, Cham, 2020. Springer International Publishing.

\bibitem{blei_latent_2003}
David~M. Blei, Andrew~Y. Ng, and Michael~I. Jordan.
\newblock Latent {Dirichlet} {Allocation}.
\newblock {\em Journal of Machine Learning Research}, 3(Jan):993--1022, 2003.

\bibitem{ebadi_understanding_2021}
Ashkan Ebadi, Pengcheng Xi, Stéphane Tremblay, Bruce Spencer, Raman Pall, and Alexander Wong.
\newblock Understanding the temporal evolution of {COVID}-19 research through machine learning and natural language processing.
\newblock {\em Scientometrics}, 126(1):725--739, January 2021.

\bibitem{raffel_exploring_2020}
Colin Raffel, Noam Shazeer, Adam Roberts, Katherine Lee, Sharan Narang, Michael Matena, Yanqi Zhou, Wei Li, and Peter~J. Liu.
\newblock Exploring the limits of transfer learning with a unified text-to-text transformer.
\newblock {\em The Journal of Machine Learning Research}, 21(1):140:5485--140:5551, January 2020.

\bibitem{ansoff_managing_1975}
H.~Igor Ansoff.
\newblock Managing {Strategic} {Surprise} by {Response} to {Weak} {Signals}.
\newblock {\em California Management Review}, 18(2):21--33, December 1975.
\newblock Publisher: SAGE Publications Inc.

\bibitem{holopainen_weak_2012}
Mari Holopainen and Marja Toivonen.
\newblock Weak signals: {Ansoff} today.
\newblock {\em Futures}, 44(3):198--205, April 2012.

\bibitem{hiltunen_future_2008}
Elina Hiltunen.
\newblock The future sign and its three dimensions.
\newblock {\em Futures}, 40(3):247--260, April 2008.

\bibitem{yoon_detecting_2012}
Janghyeok Yoon.
\newblock Detecting weak signals for long-term business opportunities using text mining of {Web} news.
\newblock {\em Expert Systems with Applications}, 39(16):12543--12550, November 2012.

\bibitem{krigsholm_applying_2019}
Pauliina Krigsholm and Kirsikka Riekkinen.
\newblock Applying {Text} {Mining} for {Identifying} {Future} {Signals} of {Land} {Administration}.
\newblock {\em Land}, 8(12):181, December 2019.
\newblock Number: 12 Publisher: Multidisciplinary Digital Publishing Institute.

\bibitem{park_analysis_2020}
Chankook Park and Seunghyun Cho.
\newblock Analysis on trends and future signs of smart grids.
\newblock {\em International Journal of Smart Grid and Clean Energy}, pages 533--543, 2020.

\bibitem{saleh_applications_2024}
Alzayat Saleh, Marcus Sheaves, Dean Jerry, and Mostafa Rahimi~Azghadi.
\newblock Applications of deep learning in fish habitat monitoring: {A} tutorial and survey.
\newblock {\em Expert Systems with Applications}, 238:121841, March 2024.

\bibitem{matley_global_2022}
Jordan~K. Matley, Natalie~V. Klinard, Ana P.~Barbosa Martins, Kim Aarestrup, Eneko Aspillaga, Steven~J. Cooke, Paul~D. Cowley, Michelle~R. Heupel, Christopher~G. Lowe, Susan~K. Lowerre-Barbieri, Hiromichi Mitamura, Jean-Sébastien Moore, Colin~A. Simpfendorfer, Michael J.~W. Stokesbury, Matthew~D. Taylor, Eva~B. Thorstad, Christopher~S. Vandergoot, and Aaron~T. Fisk.
\newblock Global trends in aquatic animal tracking with acoustic telemetry.
\newblock {\em Trends in Ecology \& Evolution}, 37(1):79--94, January 2022.
\newblock Publisher: Elsevier.

\bibitem{pandey_exploratory_2020}
Divya Pandey and Vandana Kushwaha.
\newblock An exploratory study of congestion control techniques in {Wireless} {Sensor} {Networks}.
\newblock {\em Computer Communications}, 157:257--283, May 2020.

\bibitem{makarem_design_2022}
Nabil Makarem, Wafaa Bou~Diab, Imad Mougharbel, and Naceur Malouch.
\newblock On the design of efficient congestion control for the {Constrained} {Application} {Protocol} in {IoT}.
\newblock {\em Computer Networks}, 207:108824, April 2022.

\bibitem{seo_new_2020}
Junho Seo, Sungwon Lee, Muhammad Toaha~Raza Khan, and Dongkyun Kim.
\newblock A new {CoAP} congestion control scheme considering strong and weak {RTT} for {IoUT}.
\newblock In {\em Proceedings of the 35th {Annual} {ACM} {Symposium} on {Applied} {Computing}}, {SAC} '20, pages 2158--2162, New York, NY, USA, March 2020. Association for Computing Machinery.

\bibitem{kabir_internet_2023}
Homayun Kabir, Mau-Luen Tham, and Yoong~Choon Chang.
\newblock Internet of robotic things for mobile robots: {Concepts}, technologies, challenges, applications, and future directions.
\newblock {\em Digital Communications and Networks}, 9(6):1265--1290, December 2023.

\bibitem{xie_sdrt_2010}
Peng Xie, Zhong Zhou, Zheng Peng, Jun-Hong Cui, and Zhijie Shi.
\newblock {SDRT}: {A} reliable data transport protocol for underwater sensor networks.
\newblock {\em Ad Hoc Networks}, 8(7):708--722, September 2010.

\bibitem{piermattei_cost-effective_2018}
Viviana Piermattei, Alice Madonia, Simone Bonamano, Riccardo Martellucci, Gabriele Bruzzone, Roberta Ferretti, Angelo Odetti, Maurizio Azzaro, Giuseppe Zappalà, and Marco Marcelli.
\newblock Cost-{Effective} {Technologies} to {Study} the {Arctic} {Ocean} {Environment}.
\newblock {\em Sensors}, 18(7):2257, July 2018.
\newblock Number: 7 Publisher: Multidisciplinary Digital Publishing Institute.

\bibitem{cross_innovative_2015}
J.~N. Cross, C.~W. Mordy, H.~M. Tabisola, C.~Meinig, E.~D. Cokelet, and P.~J. Stabeno.
\newblock Innovative technology development for {Arctic} {Exploration}.
\newblock In {\em {OCEANS} 2015 - {MTS}/{IEEE} {Washington}}, pages 1--8, October 2015.

\bibitem{huebert_arctic_2022}
Robert Huebert, Michael Byers, Martin La~Cour-Andersen, Rebecca Pincus, and Jonathan Quinn.
\newblock The {Arctic} as {Emerging} {Geopolitical} {Flashpoint}.
\newblock {\em Canada-United States Law Journal}, 46:32, 2022.

\bibitem{chi_prediction_2017}
Junhwa Chi and Hyun-choel Kim.
\newblock Prediction of {Arctic} {Sea} {Ice} {Concentration} {Using} a {Fully} {Data} {Driven} {Deep} {Neural} {Network}.
\newblock {\em Remote Sensing}, 9(12):1305, December 2017.
\newblock Number: 12 Publisher: Multidisciplinary Digital Publishing Institute.

\bibitem{mooradian_spectral_2014}
Greg Mooradian, Dean Richter, E.~J. Marttila, Michael Solonenko, and Edward Saade.
\newblock Spectral {Fluorescence}/{Reflectance} {Optical} {Sensor} {Systems} for {Arctic} {Oil} {Spill} {Detection} and {Mapping}.
\newblock OnePetro, February 2014.

\bibitem{kunz_deep_2008}
Clayton Kunz, Chris Murphy, Richard Camilli, Hanumant Singh, John Bailey, Ryan Eustice, Michael Jakuba, Ko-ichi Nakamura, Chris Roman, Taichi Sato, Robert~A. Sohn, and Claire Willis.
\newblock Deep sea underwater robotic exploration in the ice-covered {Arctic} ocean with {AUVs}.
\newblock In {\em 2008 {IEEE}/{RSJ} {International} {Conference} on {Intelligent} {Robots} and {Systems}}, pages 3654--3660, September 2008.
\newblock ISSN: 2153-0866.

\bibitem{su_review_2020}
Xin Su, Inam Ullah, Xiaofeng Liu, and Dongmin Choi.
\newblock A {Review} of {Underwater} {Localization} {Techniques}, {Algorithms}, and {Challenges}.
\newblock {\em Journal of Sensors}, 2020:e6403161, January 2020.
\newblock Publisher: Hindawi.

\bibitem{yan_localization_2021}
Jing Yan, Haiyan Zhao, Yuan Meng, and Xinping Guan.
\newblock {\em Localization in {Underwater} {Sensor} {Networks}}.
\newblock Wireless {Networks}. Springer Singapore, Singapore, 2021.

\bibitem{yan_auv-aided_2020}
Jing Yan, Dongbo Guo, Xiaoyuan Luo, and Xinping Guan.
\newblock {AUV}-{Aided} {Localization} for {Underwater} {Acoustic} {Sensor} {Networks} {With} {Current} {Field} {Estimation}.
\newblock {\em IEEE Transactions on Vehicular Technology}, 69(8):8855--8870, August 2020.
\newblock Conference Name: IEEE Transactions on Vehicular Technology.

\end{thebibliography}

\end{document}